%% file: main.tex
\documentclass[twocolumn]{openjournal}

\usepackage{graphicx} 

\usepackage{amsmath}
\usepackage{amssymb}
\usepackage[dvipsnames]{xcolor}
\usepackage{hyperref}
\usepackage{multirow}

\defcitealias{DESC-SRD-2018}{SRD}

\usepackage{tikz}
\newcommand{\gaussian}{
  \begin{tikzpicture}[scale=0.15]
    \draw[domain=-2.:2.,smooth,variable=\x] plot ({\x},{exp(-\x*\x)});
  \end{tikzpicture}
}

\newcommand{\tophat}{
  \begin{tikzpicture}[scale=0.15]
        \draw[thick] (-2,0) -- (-1,0) -- (-1,1) -- (1,1) -- (1,0) -- (2,0);
  \end{tikzpicture}
}

\newcommand{\kmax}{k_{\rm eff}^{\rm max}}
\newcommand{\cosmosis}{\textsc{CosmoSIS} }

\usepackage[english]{babel}
\addto\extrasenglish{%
  
}

\begin{document}
\journalinfo{The Open Journal of Astrophysics}
\submitted{submitted}

\title{Baryon-free $S_8$ tension with Stage IV cosmic shear surveys}
\input{authorlist}
\begin{abstract}
    Accurately modelling matter power spectrum effects at small scales, such as baryonic feedback, is essential to avoid significant bias in the estimation of cosmological parameters with cosmic shear. However, Stage IV surveys like LSST will be so precise that significant information can still be extracted from large scales alone. In this work, we simulate LSST Y1-like mock data and perform a cosmic shear analysis, considering different models of baryonic feedback. To focus on large scales, we apply physically motivated scale cuts which account for the redshift dependence of the multipoles in the tomographic bin. Our main focus is to study the changes in the constraining power of $S_8$ and $\Omega_{\rm m}$ parameters and assess possible effects on the tension with Planck measurements. We find that the $S_8$ tension is clearly detectable at $\kmax=0.20\,h\rm Mpc^{-1}$ in the analysis where we imposed a DES-sized tension, and at $\kmax=0.10\,h\rm Mpc^{-1}$ with a KiDS-sized tension, regardless of whether an incorrect model for baryons is assumed. However, to achieve these results, LSST will need high precision measurement of the redshift distributions, with photo-$z$ biases of the order of $10^{-3}$. Without this, the ability to constrain cosmological parameters independently of baryonic feedback - particularly regarding the $S_8$ tension - will be compromised.
\end{abstract}

\maketitle

\section{Introduction}
The most widely accepted model for describing the Universe is the $\Lambda$CDM model, which is defined by just six cosmological parameters. According to this model, ordinary baryonic matter makes up only about 5$\%$ of the Universe. Around 25$\%$ is cold dark matter (CDM), while the remaining 70$\%$ is dark energy. Although they dominate the Universe, the nature of dark matter and dark energy remains one of the biggest unresolved challenges in cosmology and particle physics.

The $\Lambda$CDM model has been supported by a series of measurements of different probes - such as anisotropies of the Cosmic Microwave Background (CMB) \cite{COBE1992-Efstathiou++, WMAP2013-Hinshaw++, Planck2018results-cosmoparam}, Baryonic Acoustic Oscillations (BAO) and Redshift Space Distortions (RDSs) \cite{2dFGalaxyRedshiftSurvey2005-Cole++, BAOinLSS2002-Eisenstein++, BOSS-RDS-GilMarin2017, DESI-IV-BAO-2024}, weak lensing of galaxies \cite{KiDS2021-Asgari++, DESY32022-cosmoconstraints-Abbott++, HSC2023-Y3Result-ShearPowSp}, and many others. However, a few questions still need an answer. 
One of the main puzzles is a discrepancy of $2\,$-$\,3\sigma$ between the measurement of the amplitude of present matter fluctuations from early-time (CMB) and late time measurements (weak lensing, galaxy clustering).
This discrepancy is often referred to as the ``$S_8$ tension'', where $S_8$ is defined as $S_8 \equiv \sigma_8 (\Omega_{\rm m}/0.3)^{0.5}$, with $\Omega_{\rm m}$ being the non-relativistic matter energy density and $\sigma_8$ representing the amplitude of matter density fluctuations.
Current literature suggests a plethora of candidate solutions for the $S_8$ tension, some of which suggest that a strong baryonic feedback on the matter power spectrum could help in solving this problem \citep{AmonEfstathiou_S8tension2022, PrestonAmonEfstathiou_S8tension2023, Preston2024-ReconstructingPk}.

While Stage III surveys - such as the Kilo Degree Survey (KiDS) \citep{KiDS-Wright2024}, the Dark Energy Survey (DES) \citep{DESY32022-cosmoconstraints-Abbott++} and the Hyper Suprime-Cam (HSC) \citep{HSC-Aihara-2022PASJ...74..247A} - are close to an end, Stage IV is starting now.
The Euclid satellite was launched in July 2023 \citep{Euclid2024-overview} and the Vera Rubin Observatory's Legacy Survey of Space and Time (LSST) \citep{DESC-SRD-2018} is scheduled to see its first light in the next year.
These experiments mark the beginning of a new era in precision cosmology, having as one of their primary focuses constraints on the amplitude of matter clustering parameter $S_8$.

On scales of a few hundred Mpc, dark matter dominates the gravitational potential and the dynamics of the Universe can be described with linear theory. In this scenario, baryons simply follow the gravitational evolution of dark matter and the Universe can be studied with gravity-only N-body simulations. On smaller scales ($k \gtrsim 0.1\,h$Mpc$^{-1}$), ordinary matter starts to dominate, redistributing the baryons leading to a suppression of the matter power spectrum in the non-linear regime \citep{Huang2018_modellingBaryons,Chisari2019-BaryonicFeedback}. Future surveys such as LSST will have increased sensitivity to these scales. This implies that baryonic effects will have a significant impact on their data, requiring a better knowledge of how these phenomena will contribute to the statistics of the large-scale structure (LSS) \citep{2006ApJ...640L.119J, Euclid:2020tff}.

The dominant physical effect of these processes, called ``baryonic feedback'', is still not well understood and, due to the highly non-linear nature of these effects across various scales, analytical modelling is impossible. As a result, they are typically extracted from hydrodynamical simulations, which rely on models with a fixed set of parameters to regulate the feedback strength. 
This approach is complementary to other works which constrain baryonic effects using different probes such as the thermal and kinetic Sunyaev-Zel'dovich effects (tSZ and kSZ) \citep{Troster2022-kSZ-baryons-KiDS, Gatti2022-kSZ-baryons-cross-corr, SchneiderGiri2022-kSZ, Bigwood2024-kSZ}.

Simulations are not the only method used to address this problem. Some works simply consider data from large-scales only, which has the clear advantage of being more model-independent. However, it comes at the cost of losing a significant part of the data \citep{Krause2021-DesY3-ModelValidation2021, DESY3-Secco2022++ModellingUncertainty}, with consequent loss of constraining power on cosmological parameters. This is more problematic for Stage-IV surveys like Euclid and LSST, because they have their highest signal-to-noise ratio (SNR) on these small, non-linear scales \citep{Euclid:2019clj}. However, since Rubin is such a precise instrument, we hope to find that even by masking out the small scales, we will still be able to effectively constrain cosmological parameters.
In this work, we explore the remaining constraining power of these large scales from LSST Y1-like data when discarding the scales affected by baryonic feedback, choosing different values of scale cuts. In particular we are interested in looking at how well the $\Omega_{\rm m}$ and $S_8$ parameters can be constrained, and the implications for detecting a tension of similar size to the existing tensions with Planck measurements.

This work is organized as follows: in \autoref{Section:BaryonicFeedback} we give a brief overview the physics of baryonic feedback and the hydrodynamical simulation models employed in our study. \autoref{Sec:Analysis} describes the methodology used in our analysis. We present our results in \autoref{Sec:Results} and conclude in \autoref{Sec:Conclusions}.

\section{Baryonic feedback and models}\label{Section:BaryonicFeedback}
\begin{figure}
    \centering
    \includegraphics[width=0.45\textwidth]{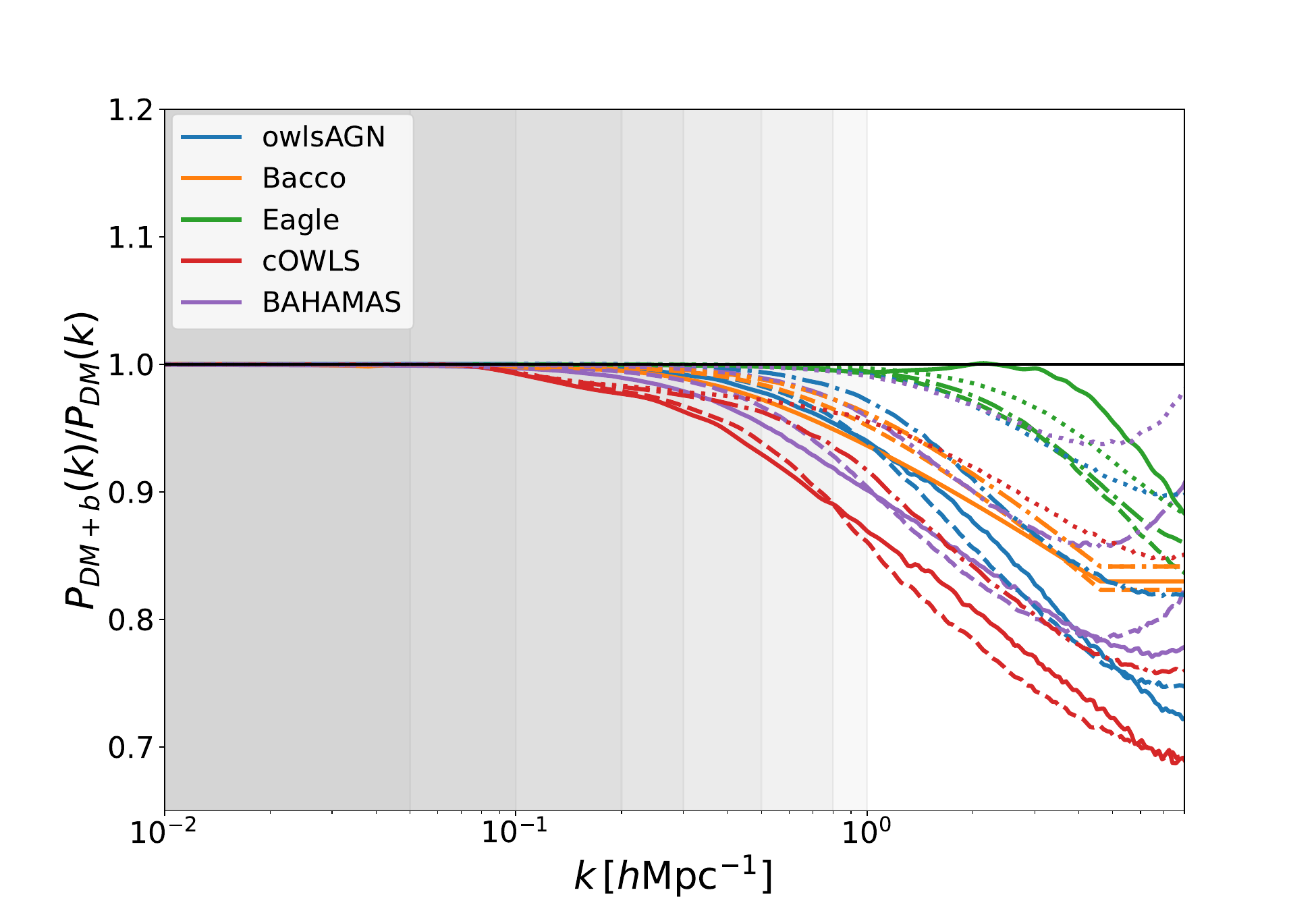}
    \caption{Comparison between dark matter-only and matter power spectrum with baryonic effects from different simulations at different redshift (solid, dashed, dash-dotted and dotted lines refer to $z=0,\, 1,\, 2,\, 3$ respectively). The grey areas represent the included range of $k$ when applying different scale cuts in the analysis. The BACCO emulator (orange lines) was evaluated with parameters as in \autoref{tab:fiducials}.} It considers the scales between $0.01$ and $5 \,h\rm Mpc^{-1}$, and is extrapolated to a flat matter power spectrum at smaller scales. The power spectra from many hydrodynamical simulations have been collected in the \href{https://powerlib.strw.leidenuniv.nl/}{Power Spectrum Library} by \citet{2020MNRAS.491.2424V}.
    \label{fig:Pk-BaryonicEffects}
\end{figure}

Modelling baryonic feedback is crucial for Stage-IV surveys, as it is required for accurate interpretation of cosmological data. Ignoring these effects can lead to greater than $5\sigma$ biases in $\Omega_{\rm m}$ and $\sigma_8$ \citep{Schneider2020-BaryonicEffects-partI} and can also lead to false detections of exotic physics \citep{Schneider2020-BaryonicEffects-partII, Tsedrik2024}.

Even without having a complete and accurate model, baryonic effects can be taken into account in different ways \citep{Huang2018_modellingBaryons}. A first way to do so is using ``baryonification'' of N-body simulations: with this method, an analytical baryon-dark matter density profile is calculated for each halo to account for baryonic effects. This profile is then added to the original gravity-only density distribution, resulting in a modified, ``baryonified'' profile, which is then used in the simulation to study the evolution of structures.
Some papers use an analytical parameterization of the effects on the matter power spectrum that can be marginalized over \citep{Schneider2016-Baryonification}, while others adopt a different parameterization of halo profiles using gas, stellar, and dark matter density components, and based on constraints from X-ray observations \citep{Schneider2019-Baryonification}. 

Another method is using the so called ``baryonic halo model'' \citep{Semboloni2011-BaryonicHaloModel}, here the Universe is modelled as matter residing in haloes which are made of dark matter, stars, and gas. This is calibrated on hydrodynamical simulations. 

Finally, the Principal Component Analysis (PCA) method, as described in \citet{Eifler-PCA}, uses both hydrodynamical and dark matter simulations to identify the modes in the observables which are more affected by baryonic effects and it marginalises over them. Thus, this method accounts for the associated systematic effects even without an explicit parameterization of the underlying physics.

\subsection{Baryonic physics} 
Many physical phenomena are included into baryonic feedback, each influencing cosmological structures and observable data across different scales.

On very small scales, often below the resolution limits of hydrodynamical simulations, the impact of stellar processes is significant. Massive stars, in particular, can inject energy and momentum into the interstellar medium through radiation, stellar winds, and supernova explosions. On larger scales, the latter two effects can contribute to the heating and movement of gas within galaxies and to the enrichment of the interstellar and intergalactic medium with metals, affecting star formation and galaxy evolution. They can also drive the expulsion of gas from galaxies, influencing the distribution of baryons on larger scales \citep{Hopkins2012MNRAS.421.3522H-feedback}.

Additionally, baryons dissipate energy through radiative processes, known as radiative cooling. This mechanism is fundamental in simulations of galaxy formation as it allows baryons to dissipate their binding energy leading to their collapse within virialized structures. As baryons condense, they exert gravitational forces that can pull in surrounding dark matter, increasing the dark matter density with a consequent enhancement of the amplitude of the matter power spectrum \citep{Rudd2008-effects-baryons-dissipation-matter-power, Semboloni2011-QuantifyingBaryons}.

Feedback from active galactic nuclei (AGN) - \textit{i.e.} accreting supermassive black holes at the centers of galaxies - is one of the strongest baryonic effects, and has a substantial impact on the matter power spectrum even on large scales. Similarly, but more strongly than supernovae, AGN feedback suppresses the power spectrum amplitude by heating and ejecting gas \citep{vanDaleen2011MNRAS.415.3649V-feedback, Velliscig2014MNRAS.442.2641V-feedback}. Models describing AGN feedback typically implement this mechanism in two modes depending on accretion efficiency. The ``quasar-mode'' occurs when the black hole is accreting efficiently, producing a hot, nuclear wind that can significantly heat and expel gas from the galaxy. On the other hand, the ``radio-'' or ``jet-mode'' is associated with lower accretion rates, where energy is injected into the surrounding medium in the form of relativistic jets, further impacting the gas dynamics and large-scale structure \citep{Sijacki2007-AGNfeedback-2modes, Rosas-Guevara2015-AGNfeedback}.

We refer the reader to \citet{ReviewCosmoSimulations-Vogelsberger2020} and references therein for a more complete review on cosmological simulations, implementation and description of baryonic feedback.

\subsection{Simulations}\label{SubSec:BaryonicEff-Simulations}
A variety of different simulation campaigns have been built to quantify and model the effect of baryons in cosmological analyses. 
In general, the feedback net effect is quantified by a boost factor, $B$, obtained by comparing the matter power spectrum from a full baryonic physics simulation with a dark matter-only scenario,
\begin{equation}
    B(k,z) = \frac{P_{\rm b+DM}(k,z)}{P_{\rm DM }(k,z)} \, ,
\end{equation}
which then is applied to the non-linear prediction of the matter power spectrum. \autoref{fig:Pk-BaryonicEffects} shows $B(k,z)$ from a suite of hydrodynamical simulations at different redshifts. 

As discussed in the previous paragraph, the enhancement of the power spectrum amplitude is due to efficient cooling of gas which eventually leads to formation of galaxies within haloes, and further concentrates the DM distribution, while the suppression at scales of a few $h\rm Mpc^{-1}$ is due to AGN feedback redistributing gas and dark matter in the simulation \citep{ Hellwig-baryonsRDS-Eagle2016, Huang2018_modellingBaryons}. However, cosmological analyses of the kind considered here do not extend to these small scales (for our work we will be interested in scales up to $k=1 \,h \rm Mpc^{-1}$).
In principle at low redshift baryons have had more time to accumulate their effect, leading to a stronger suppression; in practice this depends on the kind of effect included in the simulation, and the time at which these are switched on.

In this work, we use boost factors obtained from different large-scale structure simulations, which are briefly described below. 
Note that while these simulations were selected to cover a sufficiently wide range of feedbacks, they do not represent every possible scenario.

\begin{description}
    \item[OverWhelmingly Large Simulations (OWLS)] \citep{OWLS-Schaye_2010} is a suite of fifty simulations. By varying the implementation of the subgrid physics, it allows for the study of how different baryonic processes impact the large-scale distribution of the dark matter and baryons.
    In this work, we make use of the OWLS-AGN simulation (from now on owlsAGN), which accounts for AGN feedback and which is in good agreement with X-ray observations and can reproduce stellar masses, age distribution and formation rate as inferred by optical observations of low-redshift galaxies.

    \item[cosmo-OWLS (cOWLS)] \citep{cosmoOWLS-LeBurn2014} is a suite of simulations built to extend the OWLS project for cosmological applications. It provides a larger cosmological volume, with boxes of 400 $h^{-1}$Mpc on each side, and explores the role of baryons in affecting cluster astrophysics and non-linear structure formation. While the original OWLS simulations use the WMAP3 and Planck cosmologies, these simulations were also updated to incorporate WMAP7. In this work we used the \texttt{C-OWLS\_AGN\_Theat8.5\_WMAP7} realization \footnote{\href{https://powerlib.strw.leidenuniv.nl/}{https://powerlib.strw.leidenuniv.nl/}}.

    \item[Eagle]\citep{Eagle-Schaye2013} is a hydrodynamical simulation that assumes a flat $\Lambda$CDM cosmology with parameters from Planck2013. Although its subgrid physics is based on OWLS and cOWLS, there are significant changes in how the star formation law, the energy feedback from star formation and the accretion of gas onto black holes are implemented; they where simplified by considering different modes of stellar feedback (stellar winds, radiation pressure, supernovae) as a single sub-grid prescription. Similarly, AGN feedback (often considered as two different modes, jet and quasar modes) is considered as a unified mechanism that injects energy into the surrounding gas in a thermal form, meaning it heats up the gas without switching off radiative cooling or hydrodynamical forces.
    Furthermore, additional constraints have been applied to the distribution of galaxy sizes, which is crucial for accurately reproducing observed galaxy properties and scaling relations.

    \item[BAHAMAS] this simulation suite \citep{BahamasProject2017} has power spectra calibrated to a hydrodynamical version of the halo model, and they were specifically designed for weak lensing cross-correlation studies with the thermal Sunyaev-Zel'dovich (tSZ) effect.
    The BAHAMAS simulations builds on OWLS and cosmo-OWLS sub-grid model to create a larger suite of (400 $h^{-1}\rm Mpc^3$) boxes where the AGN feedback parameter is varied to investigate its effect on massive dark matter haloes. The most important aspect of this simulation suite, as for Eagle, is how the BAHAMAS project differs from these two sets of simulations: instead of simply exploring the effects of baryons by varying the parameters, here there is an explicit attempt to calibrate the feedback parameters to match some key observables.
    The realization used in this work is \texttt{BAHAMAS\_Theat8.0\_nu0\_WMAP9}.
    
    \item[BACCO] is a neural network-based emulator that includes baryonic effects in the non-linear matter power spectrum through a baryonification algorithm \citep{Arico2021-BACCOemulator}. It considers a $\Lambda$CDM cosmology extended with massive neutrinos and dynamical dark energy, though, in our work we fixed this to $w=-1.0$. In addition to the standard 6 cosmological parameters, BACCO has 7 additional baryonic parameters which parameterize the extent of the ejected gas ($\eta$), the density profiles of hot gas in haloes ($\theta_{\rm inn},\, M_{\rm inn},\, \theta_{\rm out}$), the fraction of gas retained in haloes of a given mass ($M_{\rm c},\, \beta$), and the characteristic halo mass scale for central galaxies ($M_{\rm 1,z0,cen}$). 
    These free parameters are fixed to the fiducial values listed in the right panel in \autoref{tab:fiducials} when creating the mock data set.
    BACCO achieves an overall precision of $\sim 1-5\%$ across its models encompassing various cosmological hydrodynamic simulations. Its range of scales is $0.01 \,h\rm Mpc^{-1}$$ < k < 5 \,h\rm Mpc^{-1}$ and it uses a constant extrapolation at smaller scales.
\end{description}

\section{Analysis}\label{Sec:Analysis}

\begin{figure}
    \centering
    \includegraphics[scale=0.33]{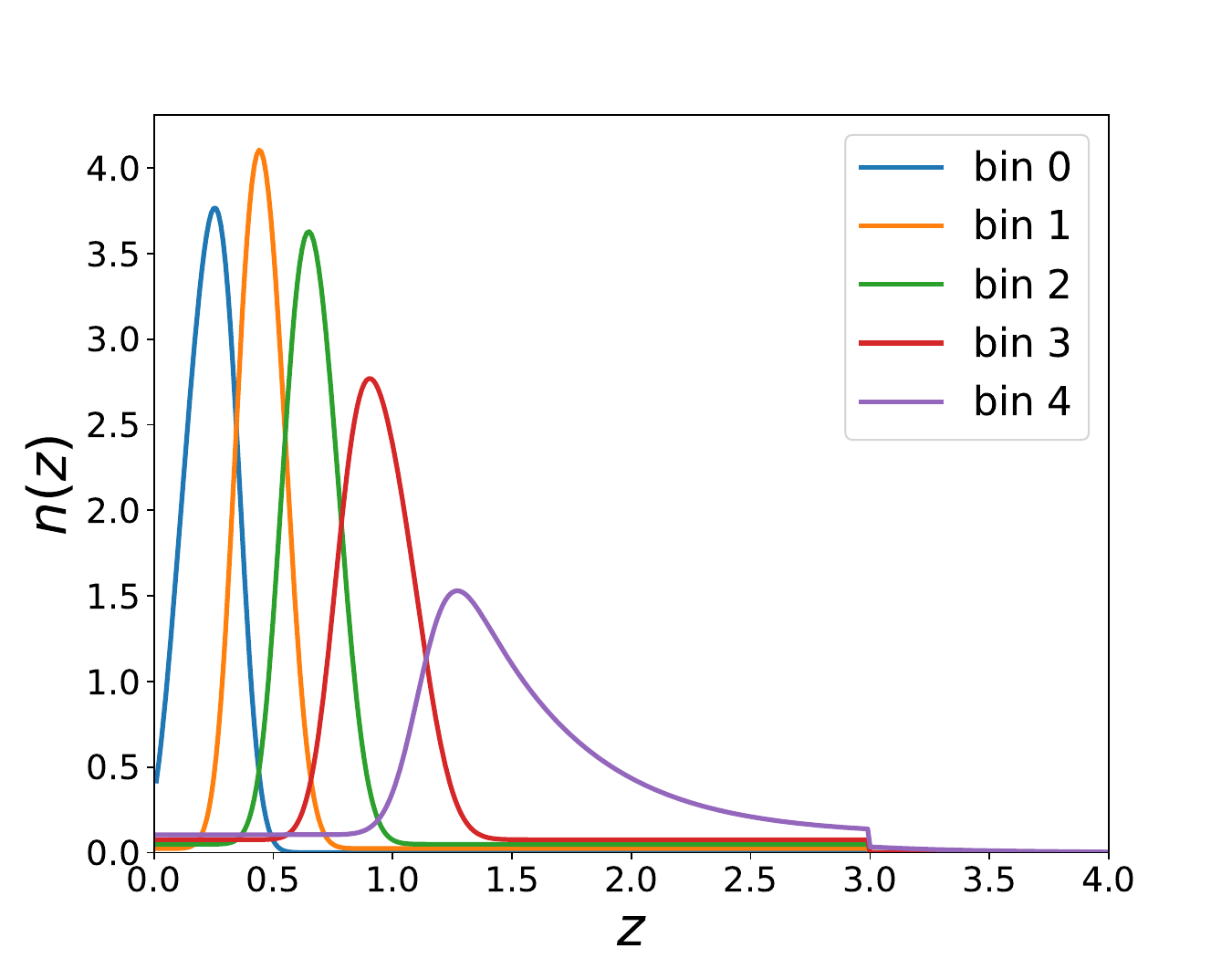} \quad \includegraphics[scale=0.29]{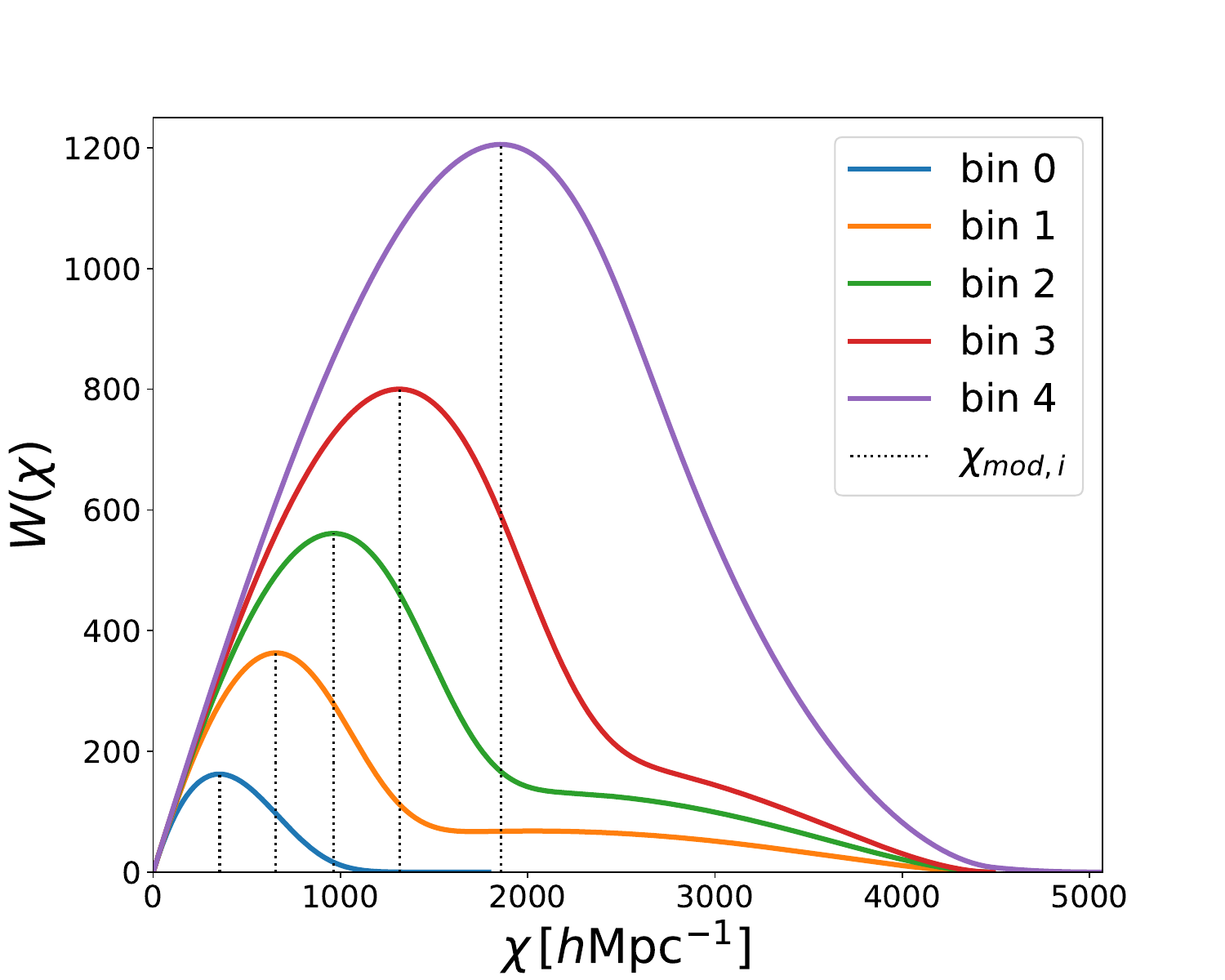}
    \caption{Redshift distribution of the tomographic bins for the source samples for the first year operation (upper panel) and the associated kernels (lower panel). The vertical dashed lines correspond to the comoving distance of the maximum value of the kernel in each bin, $\chi_{\rm mod}$.}
    \label{fig:redshift-distribution-and-kernels}
\end{figure}

\subsection{Mock data}\label{SubSec:MockData}
For this work we generated LSST-Y1 mock cosmic shear data assuming the redshift distribution shown in the top panel of \autoref{fig:redshift-distribution-and-kernels}, with five source bins in the range $z = [0.01, 4]$, based on the LSST-DESC science requirement document (hereafter \citetalias{DESC-SRD-2018}).
The data vector consists of the cosmic shear auto and cross-spectra, which are created for 20 log-spaced bins in the range $\ell=[20,3000]$ following \citet{PratZuntz-3x2ForecastRubin2023}.

Following the DESC SRD, we assume an area of $12,300 \,\mathrm{deg}^2$ for Y1
\footnote{The newest LSST configuration increases this area to $18,000 \,\mathrm{deg}^2$ \href{https://pstn-055.lsst.io/}{https://pstn-055.lsst.io/}.}
and take the same approach as \citet{PratZuntz-3x2ForecastRubin2023} using a binary mask when computing the covariance, which accounts for a reduction factor due to image defects, bright stars and so on. It should be noted that the survey footprint does not include the mask or any variation in survey depth, however, this is accounted for effectively by reducing the unmasked $n_{\rm eff}$ value by $1-f_{\rm mask}=0.12$. 

Also from the DESC SRD, the source effective number density is $n_{\rm eff, Y1}=9.78\, {\rm arcmin}^{-2}$,
while the variance on the ellipticity (the shape noise per component) is taken to be $\sigma_e = 0.26$.

The SRD redshift distribution uncertainty priors represent targets for high-precision dark energy $w_0-w_a$ measurements, rather than most-likely forecasts. As such our baseline analysis represents a rather optimistic scenario. In \autoref{sec:nz_requirements} we discuss the degradation to our results when less ambitious priors are assumed.

\subsection{Weak lensing background theory}
As previously mentioned, gravitational weak lensing is a tracer of the underlying dark matter distribution \citep{1992ApJ...388..272K, 2001PhR...340..291B}. 
In Fourier space, the matter power spectrum $P_{\delta\delta}(\Vec{k})$ is defined by the two-point correlation function
\begin{equation}
    \langle \delta(\Vec{k}) \delta(\Vec{k}') \rangle = (2\pi)^3\delta_D(\Vec{k}-\Vec{k}') P_{\delta\delta}(\Vec{k})\,.
\end{equation}

A real signal on a unit sphere, $a(\Vec{n})$, can be described in terms of its spherical harmonics coefficients $a_{\ell m}$, with $\ell\in \mathbb{N}$ and $m\in \{-\ell, \dots, \ell\}$. The angular power spectrum between two signals $a$ and $b$ is then defined as 
\begin{equation}
    \langle a_{\ell m} b_{\ell'm'} \rangle \equiv C_{ab}(\ell) \delta_{\ell \ell'} \delta_{m m'}\,,
\end{equation}
where, in our case, the signal will be the shear field $\gamma$.

Under the Limber approximation
\footnote{Please note that, although the Limber approximation will not be accurate enough for LSST \citep{Leonard2023-limber}, it will not alter the result of this work as it affects the mock data and the analysis model in the same way.}
(\citep{Limber1953, LoVerdeAfshordi-ExtendedLimberApprox}) and assuming a flat Universe cosmology, the shear power spectrum can be obtained as a projection of the 3D matter power spectrum:
\begin{equation}
     C_{\gamma\gamma}^{ij}(\ell) = \int d\chi \frac{W^i(\chi)W^j(\chi)}{\chi^2} P_{\delta\delta}\left(k=\frac{\ell+1/2}{\chi},z(\chi)\right) \, ,
    \label{eq:angular-power-sp}
\end{equation}
where the subscripts $i$ and $j$ indicate a specific redshift bin, $k$ is the wavelength number, $\ell$ is the 2D multipole moment, $\chi$ is the comoving distance, and the window function is
\begin{equation}
    W^i(\chi) = \frac{3H_0^2\Omega_{\rm m}}{2} \frac{\chi}{a(\chi)} \int_\chi^{\chi_h} d\chi^\prime  \, \frac{n^i(z)}{\bar{n}^i} \frac{dz}{d\chi^\prime} \frac{\chi^\prime - \chi}{\chi^\prime}\, ,
\end{equation}
where $H_0$ is the Hubble expansion rate today, $\Omega_{\rm m}$ is the fractional energy density of non relativistic matter, $a$ is the scale factor, $n(z)$ is the galaxy number density distribution with mean $\bar{n}$ and $\chi_h$ is the comoving distance to the horizon.

\subsubsection{Intrinsic alignments.} 
The correlation between galaxy orientation and shape is not solely determined by cosmic shear; there is an additional coherent alignment induced by the underlying tidal field on physically close galaxies, the so called intrinsic alignment (IA).
This effect generates two contamination effects for shear statistics which add to the theoretical shear angular power spectrum to give the observed $C(\ell)$ as:
\begin{equation}
    C^{\rm obs}_{ij}(\ell) = C^{\rm \gamma\gamma}_{ij}(\ell) + C^{\rm \gamma I}_{ij}(\ell) + C^{\rm II}_{ij}(\ell) \, ,
\end{equation}
where ‘I’ stands for Intrinsic and `$\gamma$' stands for shear. 
$C^{\rm \gamma\gamma}_{ij}(\ell)$, $C^{\rm \gamma I}_{ij}(\ell)$ and $C^{\rm II}_{ij}(\ell)$ are respectively linked to the matter-matter $P_{\delta\delta}(k,z)$,  matter-intrinsic $P_{\rm \delta I}(k,z)$, and the intrinsic-intrinsic power spectra $P_{\rm II}(k,z)$ as in \autoref{eq:angular-power-sp}. 

In our work, we took into account this effect in the mock data by using the non-linear alignment model (NLA-z) \citep{NLA-BridleKing2007}. According to this model, the shape distortion of a galaxy is proportional to the tidal field strength at the moment of formation. In Fourier space, the matter-intrinsic and the intrinsic-intrinsic power spectra can be written as \citep{theIAguide2024}:
\begin{align}
    P_{\rm \delta I}^{\rm NLA}(k,z) &= f_{\rm IA}(z) P_{\delta\delta}^{\rm NL}(k,z) \, , \\
    P_{\rm II}^{\rm NLA}(k,z) &= f^2_{\rm IA}(z) P_{\delta\delta}^{\rm NL}(k,z) \, ,     
\end{align}
where $P_{\delta\delta}^{\rm NL}$ is the non-linear matter power spectrum, while $f_{\rm IA}$ is defined as
\begin{equation}
    f_{\rm IA}(z) = -A_1 C_1 \rho_c \, \frac{\Omega_{\rm m}}{D(z)} \left( \frac{1+z}{1+z_{\rm piv}} \right)^{\alpha_1} \, , 
\end{equation}
where $C_1$ is a normalization constant, $\rho_c$ is the critical density today, $D(z)$ is the growth factor, and $z_{\rm piv}$ is the pivot redshift which we fix to $0.62$. The only two free parameters of the model are the dimensionless amplitude $A_1$, which determines the strength of the contamination, and $\alpha_1$ defines the redshift dependence, which takes into account the dependence of IA on the galaxy sample. Our chosen fiducial IA parameter values are taken from \citet{Fortuna2021-IntrinsicAlignment}, $A_1=0.42$ and $\alpha_1=2.21$, which are based on results from the MICE simulations for a cosmic shear survey \citep{MICEsimulation-Fosalba++2015}.

\subsection{Covariance}
The analytic model for weak lensing covariance has five contributions: three Gaussian components accounting for the shape noise, the sampling variance due to observing a finite volume, and a mix of the two; the so called super sample covariance (SSC) encodes the correlation between Fourier modes inside and modes larger than the survey window; and finally, the intra-survey non-Guassian contribution, which is subdominant and can be neglected \citep{Joachimi2021A&A...646A.129J, Euclid:2023ove}. 

Considering an area of $12,300 \, \mathrm{deg}^2$ for Y1 - as described in \autoref{SubSec:MockData} - we compute the covariance using the DESC pipeline TJPCov
\footnote{\href{https://github.com/LSSTDESC/TJPCoV}{https://github.com/LSSTDESC/TJPCoV}}. 
The Gaussian covariance is computed with the pseudo-$C_\ell$ code \texttt{NaMaster} \citep{Alonso2019-NaMaster} and the SSC is calculated using an analytic halo model.


\begin{table}
    \centering
    \begin{tabular}{|c|c|c|}
         \hline
         \multicolumn{1}{|c|}{Parameter} & \multicolumn{2}{c|}{Prior}  \\
         \hline
         $\Omega_{\rm m}$ & \(\tophat\) & $[0.24 ,0.40]$\\
         $h_0$ & \(\tophat\) & $[0.61, 0.73]$\\
         $n_{\rm s}$ & \(\tophat\) & $[0.92, 1.00]$\\
         $A_{\rm s}\times10^9$ & \(\tophat\) & $[1.7, 2.5]$\\
         $A_1$ & \(\tophat\) & $[-0.6, 1.2]$\\
         $\alpha_1$ & \(\tophat\) & $[0.1, 5.0]$\\
         $\delta_z^1$ & \(\gaussian\) & $(0.0, 0.00292)$ \\
         $\delta_z^2$ & \(\gaussian\) & $(0.0, 0.00334)$ \\ 
         $\delta_z^3$ & \(\gaussian\) & $(0.0, 0.00382)$ \\
         $\delta_z^4$ & \(\gaussian\) & $(0.0, 0.00452)$ \\
         $\delta_z^5$ & \(\gaussian\) & $(0.0, 0.00634)$ \\
         $\Omega_{\rm b}h^2$ & \(\gaussian\) & $(0.02233, 0.0004)$\\
         \hline
    \end{tabular}
    \caption{\textit{Left:} priors on the parameters sampled in the \textsc{nautilus} chain. The tophat and Gaussian symbols refer respectively to a uniform distribution $[a,b]$ between $a$ and $b$ and a Gaussian distribution $(\mu, \sigma)$ with mean $\mu$ and standard deviation $\sigma$. Priors on cosmological parameters are taken from \citet{EuclidEmulator2}. The standard deviation of the bias prior is obtained from $\delta_z^i = 0.002(1+\bar{z}_i)$ (\citetalias{DESC-SRD-2018}), where $\bar{z}_i$ indicates the mean redshift of each bin, establishing the maximum uncertainty in the mean redshift of each tomographic bin.}
    \label{tab:priors}
\end{table}

\begin{table}
    \centering
    \begin{tabular}{|c|c|c|}
         \hline
         Parameter & DES Y3 & KiDS \\
         \hline
         $\Omega_{\rm b}$ & 0.0473 & 0.0440 \\
         $\Omega_{\rm m}$ & 0.2905 & 0.2460 \\
         $h_0$ & 0.6896 & 0.7300 \\
         $\sigma_8$ & 0.7954 & 0.7773 \\
         $n_{\rm s}$ & 0.969 & 0.960 \\
         $A_{\rm s}\times10^9$ & 2.10 & 2.50 \\
         $z_{\rm piv}$ & 0.62 & 0.62 \\
         $A_1$ & 0.4 & 0.4 \\
         $\alpha_1$ & 2.2 & 2.2 \\
         $M_\nu$ & 0.06 & 0.06 \\       
         \hline
    \end{tabular}
    \quad
    \begin{tabular}{|c|c|}
         \hline
         \multicolumn{2}{|c|}{Bacco parameters} \\
         \hline
         Parameter & Fiducial \\
         \hline
         $\text{log}\, M_c$ & 14.0 \\
         $\text{log}\, \eta$ & -0.3 \\
         $\text{log}\, \beta$ & -0.22 \\
         $\text{log}\, M_{\rm 1,z0,cen}$ & 10.5 \\
         $\text{log}\, \theta_{\rm out}$ & 0.25 \\
         $\text{log}\, \theta_{\rm inn}$ & -0.86 \\
         $\text{log}\, M_{\rm inn}$ &13.4 \\
         \hline
    \end{tabular}
    \caption{\textit{Left:} fiducial values for all parameters, the cosmological parameters following a DES Y3-like \citep{DESY3-Secco2022++ModellingUncertainty} and a KiDS-like survey \citep{KiDS2021-Asgari++}. \textit{Right:} fiducial values for the baryonic parameters; used only for Bacco.}
    \label{tab:fiducials}
\end{table}

\begin{table*}
    \centering
    \begin{tabular}{|c|c|c||c|c|c|c|c|c|c|c|}
         \hline
          \multirow{2}{*}{$z$-bin} & \multirow{2}{*}{$z_{\rm med}$} & \multirow{2}{*}{$z_{\rm mod}$} & \multirow{2}{*}{} &  \multicolumn{7}{c|}{$\kmax \,\, [h \rm Mpc^{-1}]$}  \\ [1ex]
         \cline{5-11}
         & & & & 0.05 & 0.10 & 0.20 & 0.30 & 0.50 & 0.80 & 1.00 \\
         \hline
         0 & 0.46 & 0.12 & $\ell_{\rm max}^{0,j}$ & 23 & 47  & 94  & 142 & 263 & 377  & 472  \\
         1 & 0.67 & 0.23 & $\ell_{\rm max}^{1,j}$ & 39 & 79  & 158 & 238 & 396 & 633  & 792  \\
         2 & 0.91 & 0.36 & $\ell_{\rm max}^{2,j}$ & 53 & 106 & 212 & 319 & 531 & 850  & 1062 \\
         3 & 1.26 & 0.50 & $\ell_{\rm max}^{3,j}$ & 65 & 130 & 261 & 391 & 652 & 1043 & 1304 \\
         4 & 2.17 & 0.77 & $\ell_{\rm max}^{4,j}$ & 78 & 157 & 313 & 471 & 785 & 1255 & 1569 \\
         \hline
    \end{tabular}
    \caption{From the left: number or redshift bin, redshift corresponding to the median of $n(z)$ ($z_{\rm med}$), redshift corresponding to the maximum value of the kernel ($z_{\rm mod}$), and scale cuts in $\ell$ corresponding to each chosen $\kmax$ in all redshift bins. The specific bins combinations for $\ell{\rm max}$ are reported with $(i,j)$, where $i$ and $j$ indicate the first bin (column) and the second bin (row), respectively.}
    \label{tab:scalecuts-k-ell}
\end{table*}


\subsection{\cosmosis settings}
The main tool we use in this work is the latest version of the publicly available software for cosmological parameter estimation \cosmosis
\footnote{\href{https://github.com/joezuntz/cosmosis}{https://github.com/joezuntz/cosmosis}}.

The main steps of our model are as follows. Perturbations in the primordial Universe and the linear matter power spectra are computed with \texttt{CAMB}
\footnote{\href{https://camb.info/}{https://camb.info/}}
\citep{CAMB-Lewis_2000}, while the non-linear corrections are added with the \texttt{EuclidEmulator2} module \citep{EuclidEmulator2}, which calculates a boost factor to scale the linear to the non-linear power spectrum. We also consider the intrinsic alignment and baryonic effect contributions. Specifically, we use the \cosmosis implementations of the baryonic boost factor of the models listed in \autoref{SubSec:BaryonicEff-Simulations}, \textit{i.e.} owlsAGN, cOWLS, Eagle, BAHAMAS simulations, and the BACCO emulator.

The \cosmosis pipeline is used to produce the initial mock data set and the subsequent analysis is performed with \textsc{nautilus} \citep{Nautilus-Lange2023}, an importance nested sampling method which uses deep learning to boost the sampling efficiency. 
To ensure convergence, we choose to sample the posterior using $2\times10^3$ live points.

With this sampler, given the priors and the fiducial values of the parameters as listed in Tables \ref{tab:priors} and \ref{tab:fiducials}, the code provides the posterior distribution of six cosmological parameters ($h_0$, $\Omega_{\rm b} h^2$, $\Omega_{\rm m}$, $n_{\rm s}$, $A_{\rm s}$) - with $\sigma_8$ and $S_8$ derived but not sampled - along with two IA parameters ($A_1$, $\alpha_1$) and the five photo-$z$ bias $\delta_z^i$ defined as $n^i_{\rm observed}(z) = n^i_{\rm true}(z - \delta_z^i)$, one for each redshift bin $i$. In this work we focus on the two parameters $S_8$ and $\Omega_{\rm m}$, so we marginalize over all the others for most plots and metrics. 
Note that when using BACCO for modelling the baryons we also vary the baryonic parameters ($M_{\rm c}$, $\beta$, $\theta_{\rm inn}$, $\theta_{\rm out}$, $M_{\rm inn}$).

It is also crucial to observe that because \texttt{EuclidEmulator2} is built from N-body simulations, it is essential to stick to the parameter ranges it was trained on, hence our choice for the cosmological parameter priors.

\subsection{Choice of fiducial values}
We consider two different set of fiducial values for cosmological parameters close to the recent results of stage III surveys, so that we impose a DES Y3-sized and a KiDS-sized tension with Planck measurements. 

In the first case we used a value of the matter perturbation amplitude as in DES Y3 results \citep{DESY3-Secco2022++ModellingUncertainty}, as well as the other cosmological parameters.
For the second set of fiducials we took the values to be similar to KiDS results from a COSEBIs (Complete Orthogonal Sets of E/B-Integrals) analysis \citep{KiDS2021-Asgari++}. It should be noted, however, that to remain within the parameter range of \texttt{EuclidEmulator2}, we selected a different value of $\sigma_8$, as the original $A_{\rm s}$ value would have otherwise been outside the allowed prior range.

\subsection{Scale cuts}
As previously mentioned, modelling non-linear effects, particularly baryonic effects, is critically important for next-generation surveys.
However, given the precision of upcoming surveys, it maybe still relevant to temporarily avoid this issue by either reducing sensitivity to small-scale data (see \citet{DESY1-Huang2021++ConstrainingBaryonicPhysics, maraio2024mitigatingbaryonfeedbackbias} and references therein) or by applying stringent scale cuts to exclude small scales where baryonic physics is relevant.

The latter approach has the advantage of being model-independent and may help avoid biases in inferred models and cosmological parameters. Nevertheless, this method inevitably discards a substantial portion of the data, resulting in significantly reduced constraining power on cosmological parameter estimation - see for example \citet{Chisari2019-BaryonicFeedback}. 
That being said, the primary focus of this work is to understand how much we can learn from LSST Y1 data once we discard the part of the data vector most heavily contaminated by baryons.

\subsubsection{Choice of scale cuts}
The choice of scale cuts is not unique: some studies adopt a single value of $\ell_{\rm max}$ for all redshift bins \citep{,Schneider2020-BaryonicEffects-partI, GarciaGarcia2024}, others use optimized $\Lambda$CDM angular scale cuts \citep{DESY3-Secco2022++ModellingUncertainty}, or remove scales from the data vector to ensure baryon contamination remains within a target fraction at any physical scale \citep{DESY1-Troxel2018++CosmologicalConstraints}. 
In this work, we opt for a more physical description of the scale cuts, taking into account their dependence on redshift - \textit{ie.} considering that the same scale at different redshifts corresponds to different angular distances. Additionally, we incorporate the lensing efficiency and the number of sources through the weighting function.

We use the notation $\kmax$ to denote a region in $k$-space which we wish to retain. This corresponds to different values of $\ell_{\rm max}$ for each tomographic bin, where the data is included if $\ell<\ell_{\rm max}$. Once a value of $\kmax$ is chosen, the corresponding $\ell_{\rm max}$ can be found as \citep{Krause2021-DesY3-ModelValidation2021}
\begin{align}
    \theta = \frac{\pi}{\kmax d} \,\,\, \Rightarrow \,\,\, \ell_{\rm max} = \frac{\pi}{\theta} = \kmax d \, ,
    \label{scalecuts-formula}
\end{align}
where $\theta$ is the physical angular distance and $d$ is the angular diameter distance $d = \chi_{\rm mod}/(1 + z_{\rm mod})$. Here $z_{\rm mod}$ and $\chi_{\rm mod}$ are the redshift of the maximum value of the kernel $W(\chi)$ and the corresponding comoving distance ($\chi_{\rm mod} = \chi(z_{\rm mod})$); see the lower panel in \autoref{fig:redshift-distribution-and-kernels}. 
Note that this approach differs from other works that consider $z_{\rm mod}$ to be the redshift of the mean (or median) $n(z)$ value within each bin. Our choice accounts for the offset between $n(z)$ and $W(\chi)$. More importantly, it is a physically motivated choice, as lensing is most sensitive at the distance corresponding to the peak value of the kernel.

While the minimum $\ell$ is kept fixed at 20, in this work we investigate scale cuts corresponding to $\kmax=0.05$, $0.10$, $0.20$, $0.30$, $0.50$, $0.80$ and $1.00\,h\rm Mpc^{-1}$. The corresponding cuts in $\ell$ are reported in \autoref{tab:scalecuts-k-ell}, together with the value of the redshift corresponding to the maximum value of the kernel and the median of the density distribution.

\subsection{Tension metrics}    
A variety of different methods to translate two data set constraints into a probabilistic measure of tension between them exists in the literature, and the choice of which method to use depends on several factors. A comprehensive review can be found in \citet{Lemos_TensionMetrics_PlanckDES2021}. Following their argument, we use ``parameter difference'' as a tension estimator in our work \citep{RaveriHu-2019-parametershift,  RaveriZacharegkasHu-2020-parametershift, RaveriDoux-2021}

According to this method, the parameter difference probability density of two uncorrelated data sets $A$ and $B$, is simply the convolution
\begin{equation}
    \mathcal{P}(\Delta\theta) = \int_{V_p} \mathcal{P}_A(\theta)\mathcal{P}_{B}(\theta-\Delta\theta) d\theta\, ,
\end{equation}
where $V_p$ is the volume of non vanishing prior, and $\mathcal{P}_A$ and $\mathcal{P}_B$ are the posterior distributions of the two parameters. The probability that an actual shift exists is obtained from
\begin{equation}
    \Delta = \int_{\mathcal{P}(\Delta\theta)>\mathcal{P}(0)} \mathcal{P}(\Delta\theta) d\Delta\theta \, ,
    \label{parameter-shift-Delta}
\end{equation}
where the volume of integration is defined as the region of parameter space above the isocontour of no shift $\Delta\theta=0$.

To obtain the effective number of sigma, $N_{\sigma,\rm eff}$, that corresponds to the measured shift we use the \texttt{tensiometer} code by \citet{RaveriHu-2019-parametershift}
\footnote{\href{https://tensiometer.readthedocs.io/en/latest/}{https://tensiometer.readthedocs.io/en/latest/}}
, which computes $\Delta$ and $N_{\sigma}$:
\begin{equation}
    N_{\sigma} = \sqrt{2} \, \text{erf}^{-1}(\Delta) \, .
\end{equation}

Note that, by assuming DES Y3-like (and later KiDS-like) fiducial values we are explicitly introducing a tension between our mock data and the Planck measurements. Specifically, the measured values of $S_8$ by the three experiments are:
\begin{align*}
    & S_8 = 0.834^{+0.016}_{-0.016} \,\,\,\,\, \rm Planck2018-TTTEEE\\
    & S_8 = 0.772^{+0.018}_{-0.017} \,\,\,\,\, \rm DES\,Y3 \\
    & S_8 = 0.759^{+0.024}_{-0.021} \,\,\,\,\, \rm KiDS-1000 \, .
\end{align*}

In our analysis, this ``true tension'' is evaluated between Planck results and the trivial run in which BAHAMAS simulations were utilized both for constructing the mock data and for their subsequent analysis.

\subsection{Angular power spectrum}
\begin{figure*}
    \centering
    \includegraphics[width=0.65\linewidth]{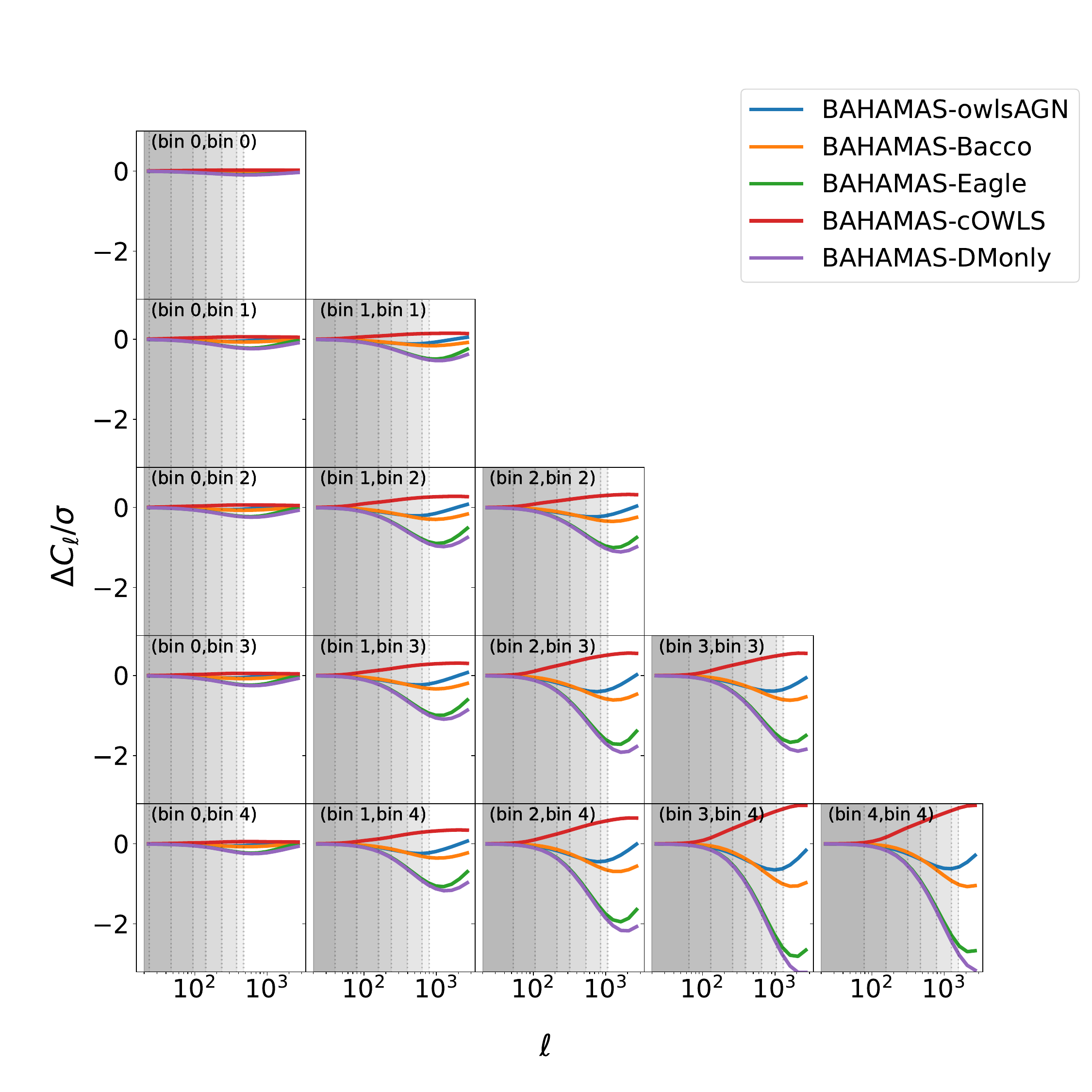}
    \caption{$\Delta C_\ell/\sigma$ for different combinations of baryonic models defined as in (\autoref{eq:deltaCls}). The grey areas represent the $\ell<\ell_{\rm max}$ area.}
    \label{fig:deltaCls}
\end{figure*}

As the actual observable is the angular power spectrum, we are interested in its behaviour at high $\ell$ (where the baryonic feedback should be dominant) and in different redshift bins. In particular, considering BAHAMAS as a reference baryonic model, we use the \cosmosis \texttt{test} sampler to compute the following quantity:
\begin{equation}
    \frac{\Delta C_\ell}{\sigma} = \frac{C_\ell^{\rm reference} - C_\ell^{\rm alternative}}{\sigma^{\rm reference}} \, ,
    \label{eq:deltaCls}
\end{equation}
where the superscript ``alternative'' indicates all other baryonic models we considered in our analysis. In \autoref{fig:deltaCls} we present the resulting $\Delta C_\ell/\sigma$ in all tomographic bins. In the majority of cases, at high redshift the effect of the baryons should be lower
\footnote{Though see the comment in \autoref{SubSec:BaryonicEff-Simulations}.
}
(see \autoref{fig:Pk-BaryonicEffects}), thus one might expect a consistent behaviour of $\Delta C_\ell/\sigma$ at higher $\ell$ within each bin. However, the more a source is distant, the more it is lensed, and the noise is then lower - \textit{i.e.} increasing $|\Delta C_\ell/\sigma|$. Since the difference between the various cases is increasing at higher redshift bins, the second effect is dominant.

\section{Results}\label{Sec:Results}
\begin{table}
    \small  
    \centering
    \begin{tabular}{|c|c|}
        \hline
         Mock data & Fitted model \\
        \hline
         BAHAMAS & BAHAMAS \\
         BAHAMAS & owlsAGN \\
         BAHAMAS & cOWLS \\
         BAHAMAS & Eagle \\
         BAHAMAS & Bacco \\
         Bacco & BAHAMAS\\
         owlsAGN & BAHAMAS\\
         BAHAMAS & DMonly \\
         DMonly & DMonly \\
        \hline
    \end{tabular}
    \caption{Combinations of different baryonic models considered for creating the mock data (left column) and for the analysis model (right column).}
    \label{tab:different-model-combinations}
    \small
\end{table}

In \autoref{Section:BaryonicFeedback} we have seen how different baryonic models affect the matter power spectrum at small scales and how their effect can in principle be neglected simply by masking those scales. 
We are particularly interested in the possibility of finding a scale cut value at which baryonic effects become negligible, such that the choice of model becomes irrelevant, and see if with this scale the constraining power is still sufficiently strong to detect KiDS and DES-sized tensions.
To this end, we consider the model combinations listed in \autoref{tab:different-model-combinations}, for which we run \cosmosis with the \textsc{nautilus} sampler across various scale cut values. 
Notice that the last case in the table is not physical, but is a``worst case scenario'' - \textit{ie.} assuming to use a dark matter-only model to analyse data containing baryonic effects. In order to check the pipeline, we also perform a dark matter-only run (no baryons in the data or in the model), and one with the same baryonic model (BAHAMAS) for the data and the model.

For brevity, we will indicate the runs as \textit{model1-model2}, where \textit{model1} is the baryonic model used to create the mock data and \textit{model2} the one used in the analysis to fit the mock data (\textit{eg.} BAHAMAS-cOWLS).

\subsection{Results overview}
We report here our main results - which will be explained in more detail in the following sections - focusing on the loss of constraining power due to the applied scale cuts and its effect on our data, specifically in relation to the Planck results (analysing the $S_8$ tension) \citep{Planck2018results-cosmoparam}.

\begin{figure*}
    \centering
    \includegraphics[scale=0.4]{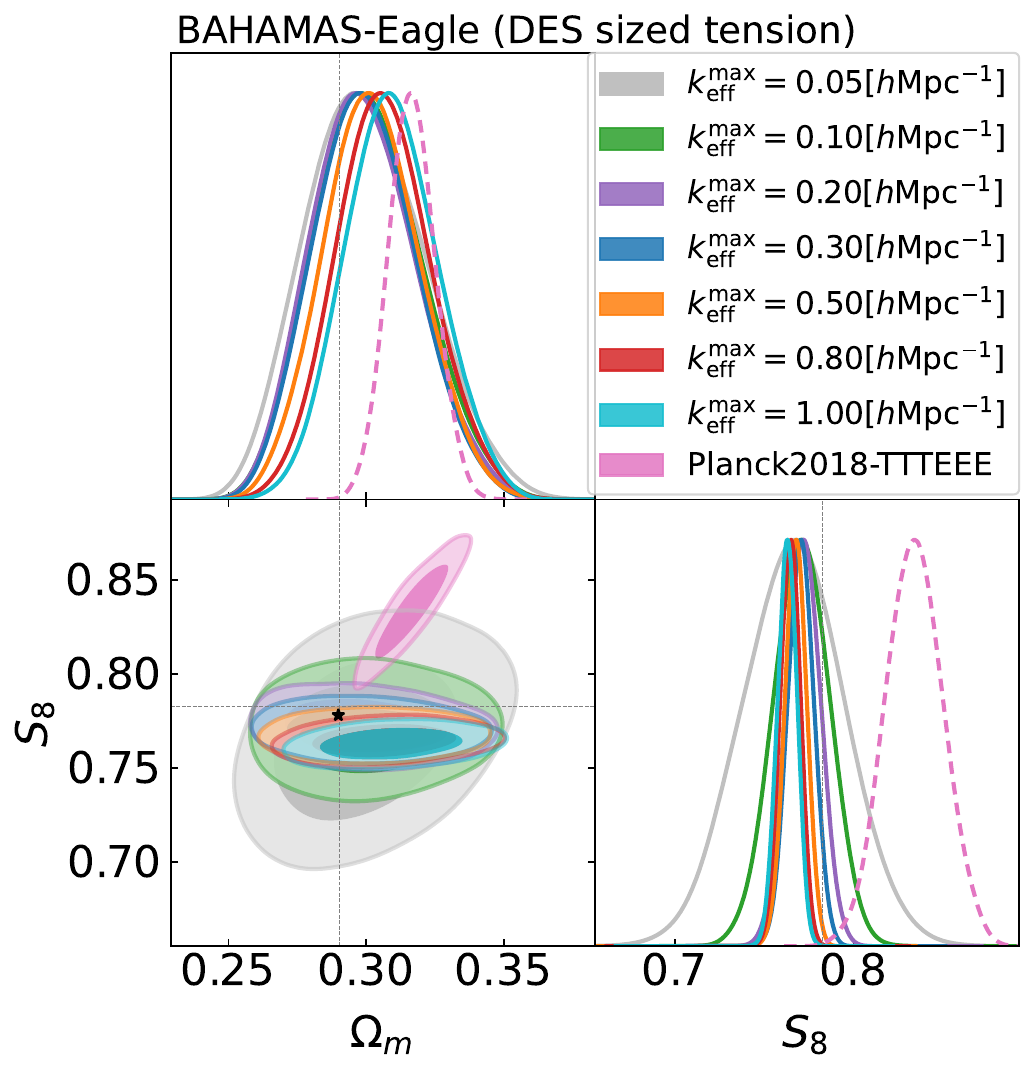} \quad
    \includegraphics[scale=0.4]{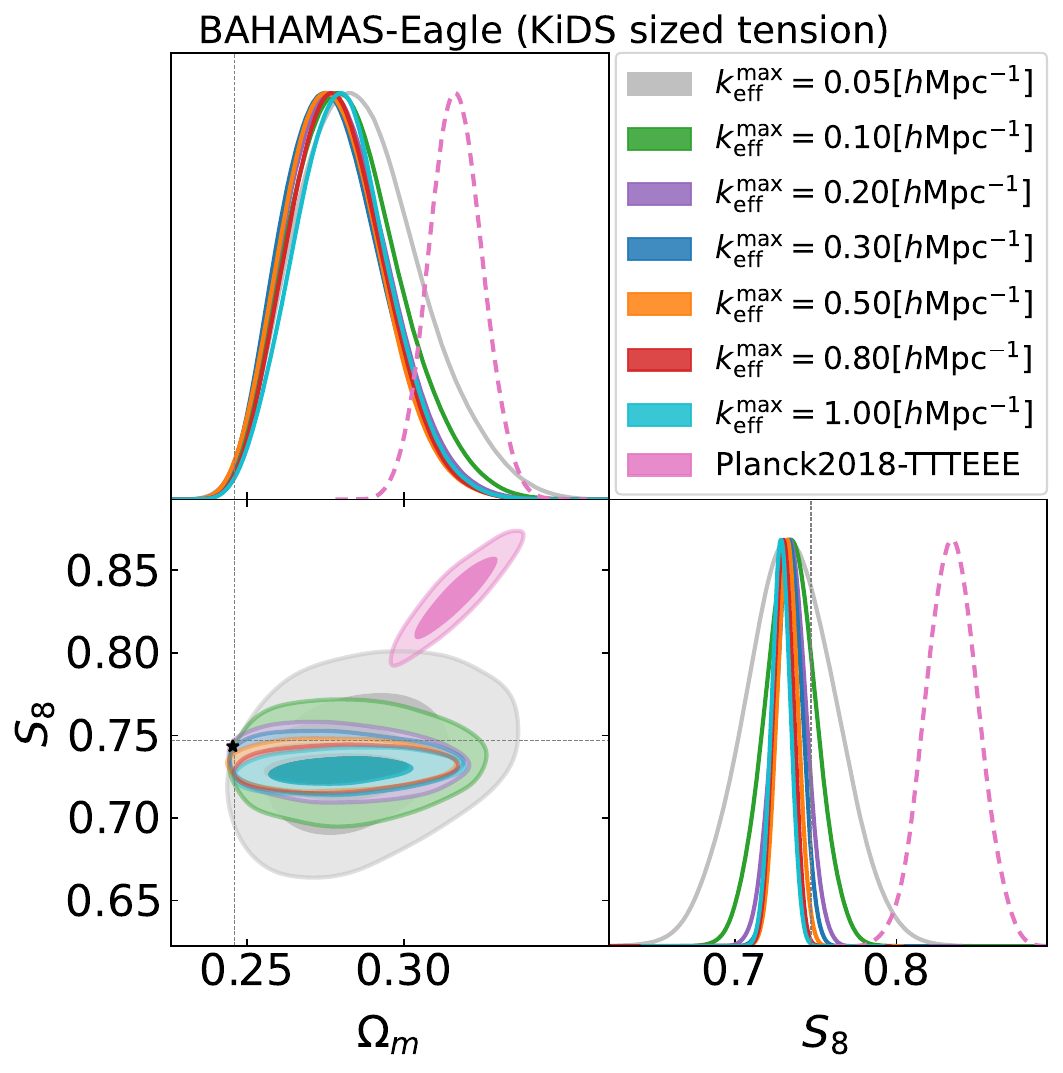} \quad 
    \includegraphics[scale=0.4]{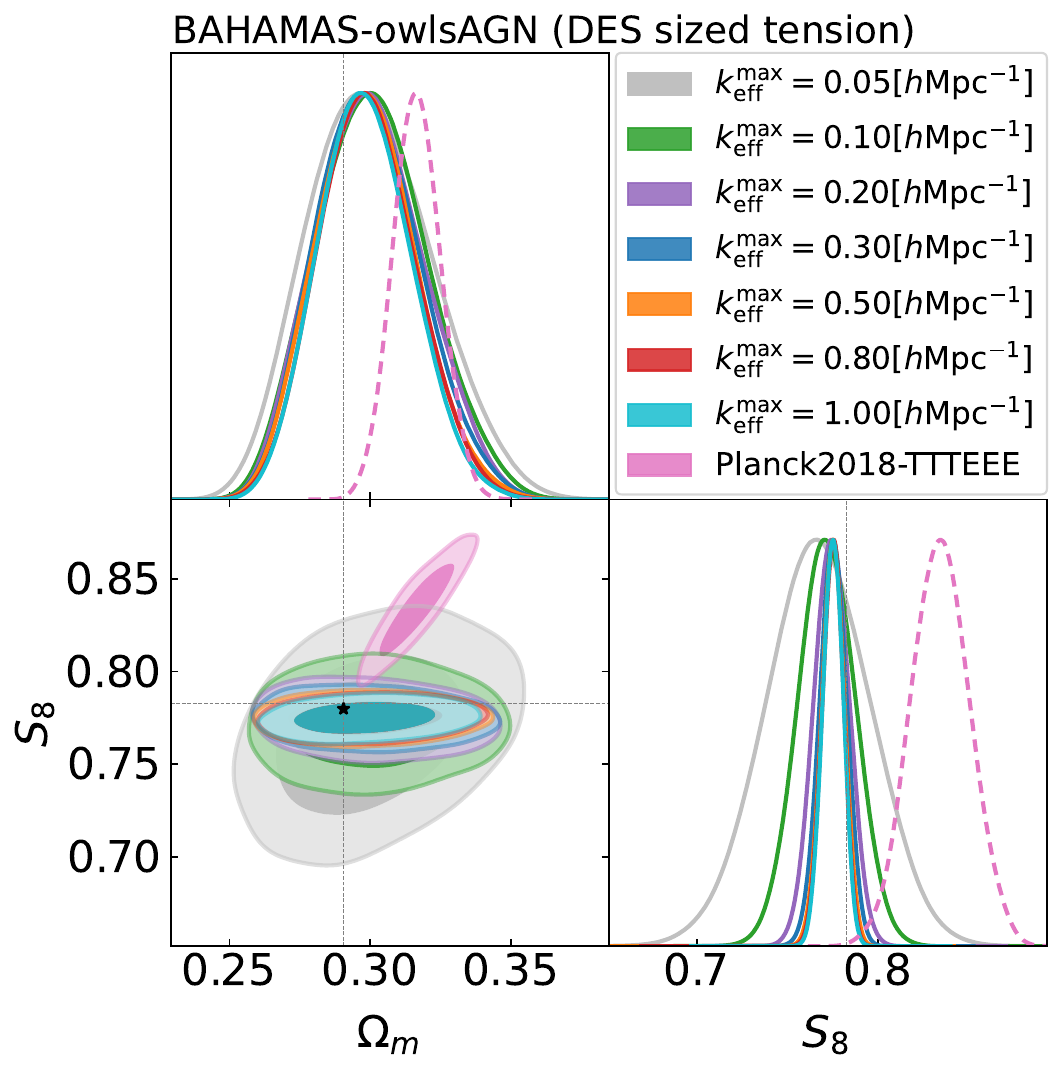} \quad \includegraphics[scale=0.4]{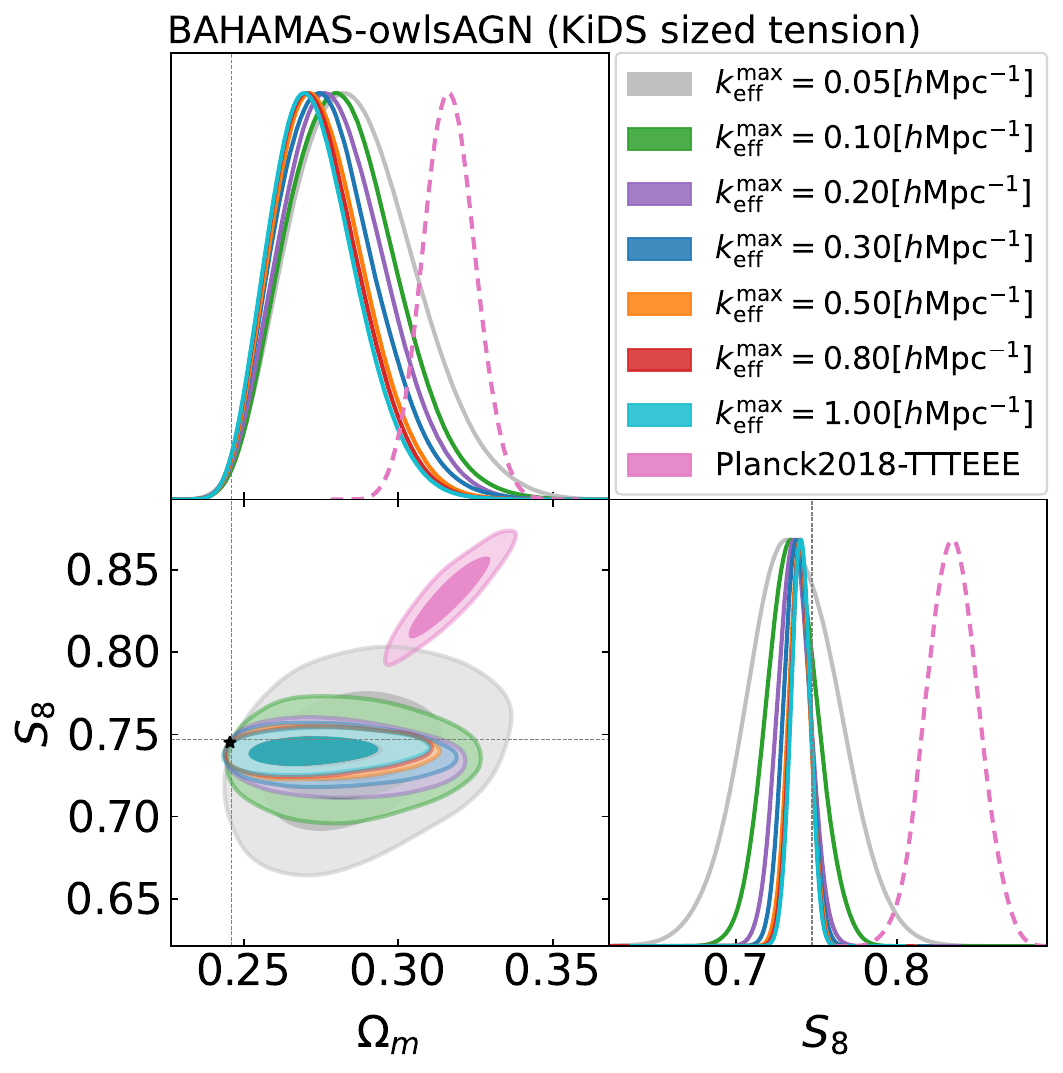}
    \includegraphics[scale=0.4]{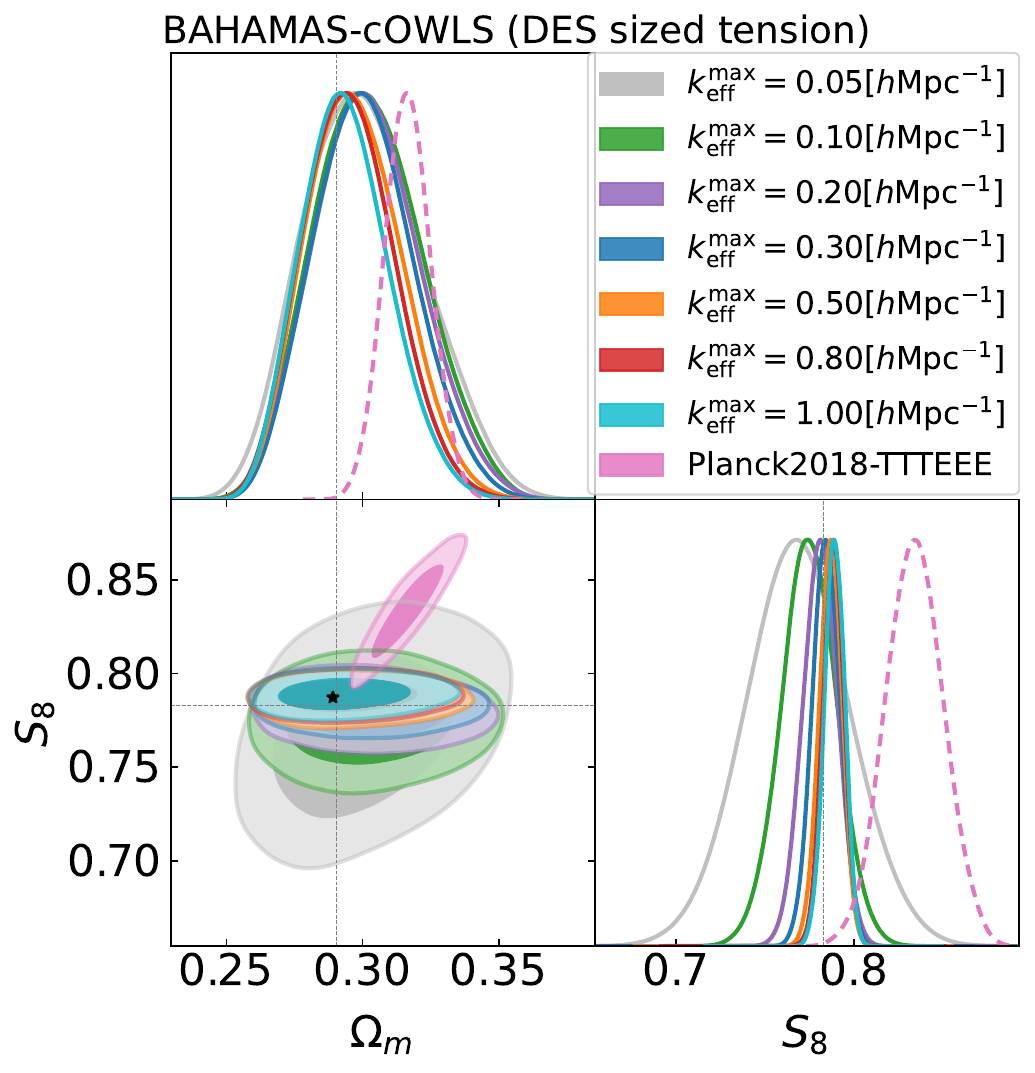} \quad
    \includegraphics[scale=0.4]{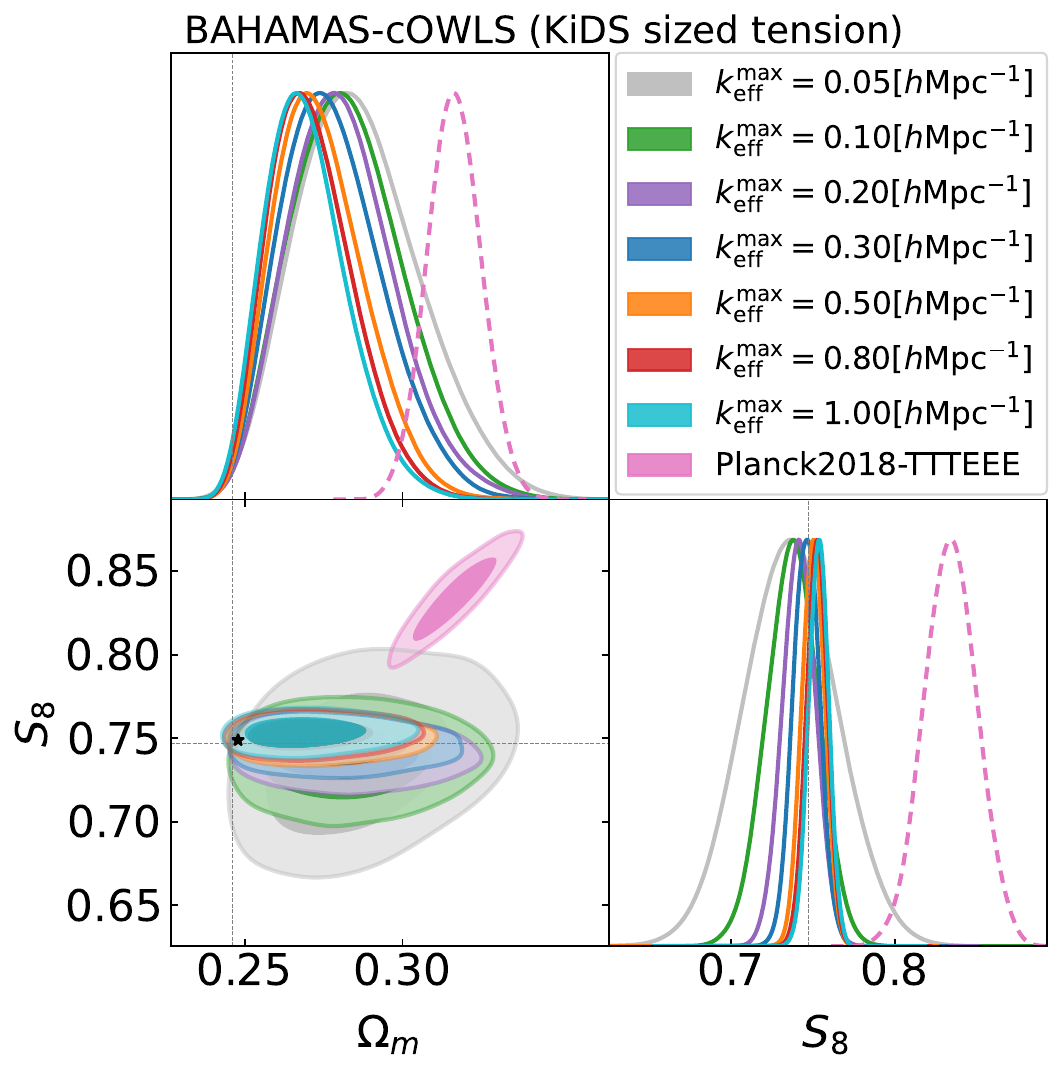} 
    \caption{The $68\%$ and $95\%$ 2D confidence regions and 1D posteriors for $\Omega_{\rm m}$ and $S_8$ compared to the Planck contours (pink ellipses) for different baryonic model combinations. Mock data created with a DES Y3-like (left) and with KiDS like tension (right). Different choices of scale cuts, $\kmax$, are reported with different colours. The fiducial values used creating the mock data (see \autoref{tab:fiducials}) are shown with the dashed grey lines. The black ${*}$ symbols show the maximum a posteriori (MAP) estimate at $\kmax=0.10 \,h \rm Mpc^{-1}$ - which confirms that our sampling is accurate, and the offset of the ellipses from the fiducial values is just a boundary effect caused by our choice of $\Omega_{\rm m}$ being near the edge of the prior range defined by \texttt{EuclidEmulator2}.}
    \label{fig:corner-plots-Omegam-S8-DES-KiDS}
\end{figure*}

\begin{figure*}
    \centering
    \includegraphics[scale=0.6]{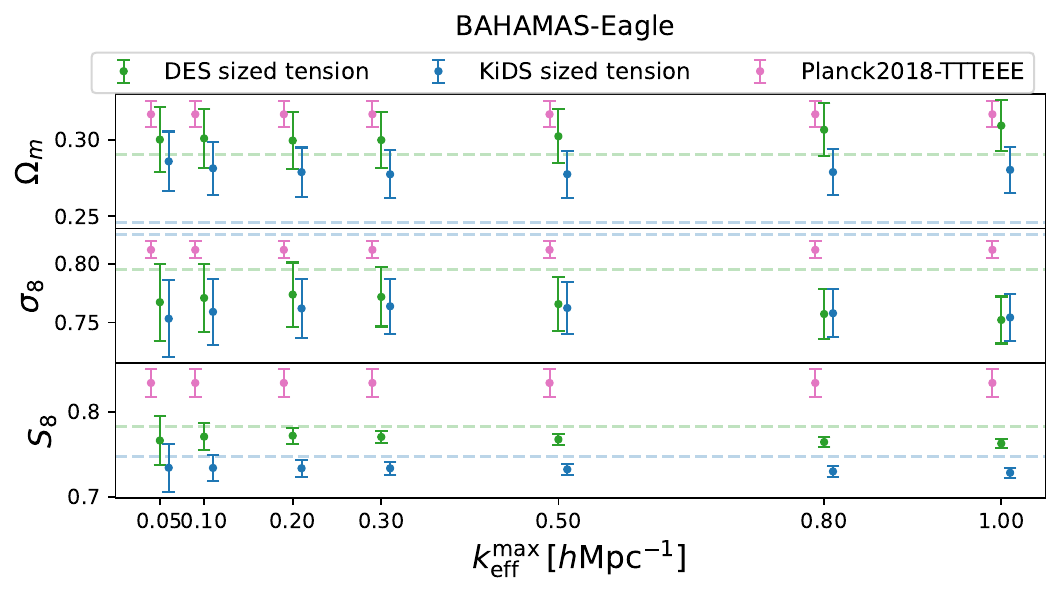} \quad 
    \includegraphics[scale=0.6]{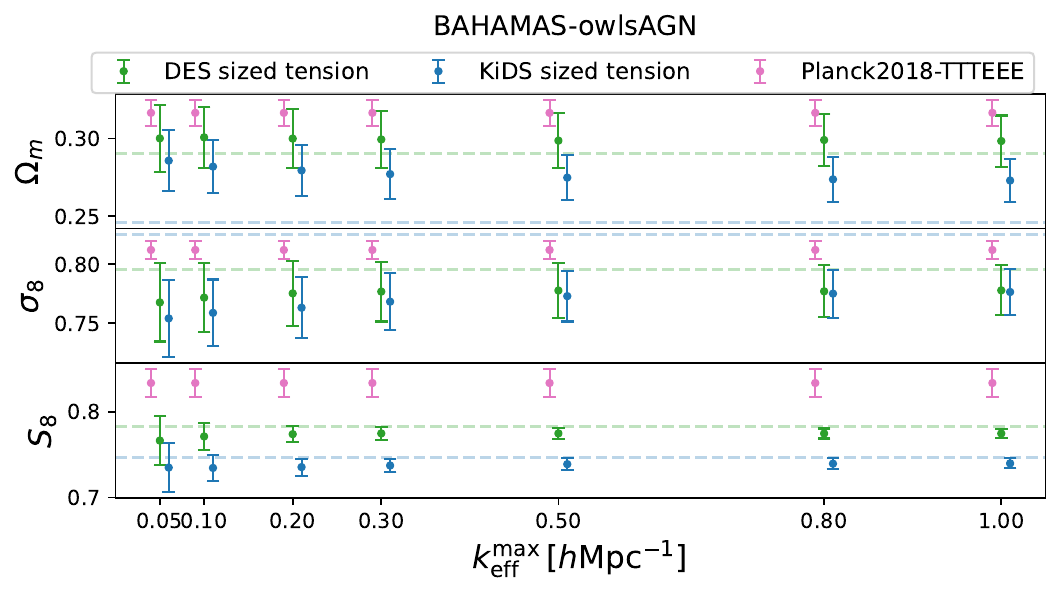} \quad 
    \includegraphics[scale=0.6]{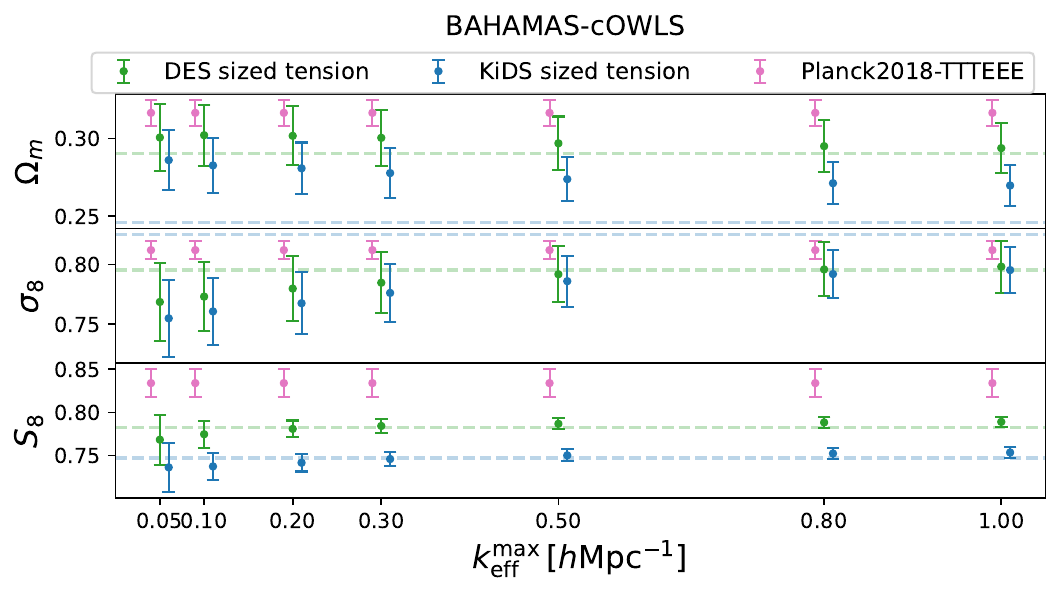} 
    \caption{Comparison of the mean and standard deviation of $\Omega_{\rm m}$, $\sigma_8$ and $S_8$ from the analysis with DES Y3-like mock data (green) and KiDS-like (blue) with respect to Planck results (pink) at each scale cut. The dashed lines correspond to the fiducial values used to create the mock data. \textit{Top:} BAHAMAS-Eagle, \textit{center:} BAHAMAS-owlsAGN \textit{bottom:} BAHAMAS-cOWLS.}
    \label{fig:errorbars_DES_KiDS_Planck}
\end{figure*}

\begin{figure*}
    \centering
    \includegraphics[width=\textwidth]{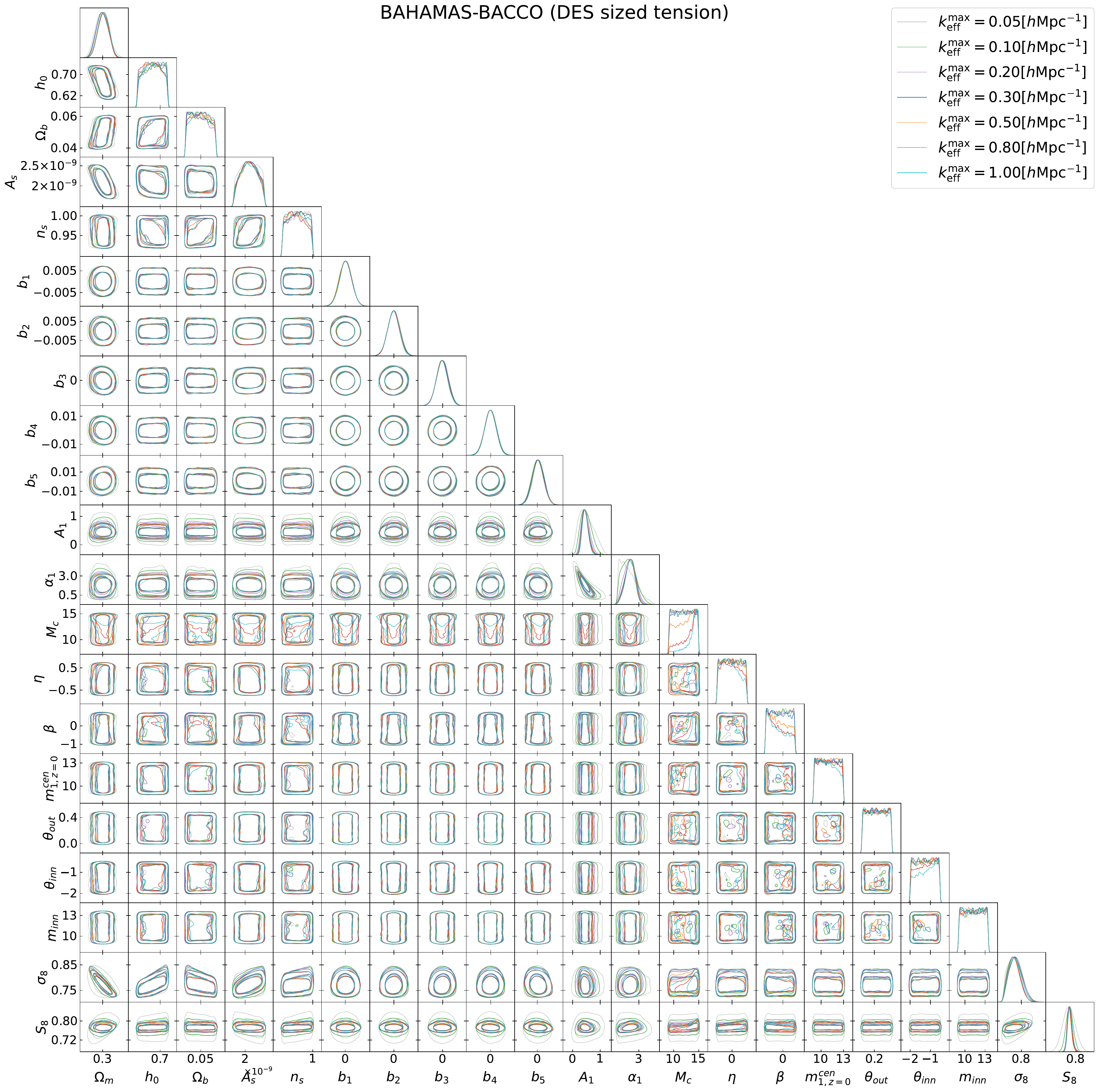}
    \caption{Contour plot for all parameters for the case BAHAMAS-Bacco. }
    \label{fig:cornerplot-all-params-bacco}
\end{figure*}

\autoref{fig:corner-plots-Omegam-S8-DES-KiDS} presents examples of contour plots illustrating the results of our analysis compared to the Planck measurements. Specifically, we choose to report the cases with Eagle, cOWLS and owlsAGN simulations in the analysis model as they are, respectively, the weakest, the strongest, and the most similar models with respect to the BAHAMAS simulation, which was used to create the mock data. Additionally, \autoref{tab:avg_std_powerloss} shows the mean and standard deviation of $\Omega_{\rm m}$, $\sigma_8$ and $S_8$ parameters for all the chosen scale cut values, and \autoref{fig:errorbars_DES_KiDS_Planck} compares these values to the Planck results. 
The complete contour plots for two example analyses are reported in \autoref{fig:cornerplot-all-params} and \autoref{fig:cornerplot-all-params-bacco}.

Two main results are evident: baryons affect the data when considering $\kmax\gtrsim0.2$, but even with this aggressive scale cut we can clearly distinguish between the LSST Y1 measurements and those of Planck. Furthermore, in the analysis using KiDS fiducial data, the tension remains visible even with extreme scale cuts.
However, note that these results are achievable only with sufficiently small errors on the photometric redshift distribution $n(z)$ (of the order $10^{-3}$, as in \autoref{tab:priors}).

\subsection{Constraints on cosmological parameters}\label{SubSec:Constrains}
The loss of constraining power due to scale cuts concerns predominantly the $S_8$ (and the $\sigma_8$) parameter. In particular, while the error on $\Omega_{\rm m}$ slightly changes when considering just large scales, with a maximum variation between $\kmax = 0.05$ and $1.00\, h \rm Mpc^{-1}$ of $27\%$, the error on $S_8$ increases of almost by a factor of 5 in the $1-\sigma$ level of the marginalized posterior; however, disregarding the severe cut at $k=0.05\, h \rm Mpc^{-1}$, the error on $S_8$ changes only of a factor of 2.6. 

Also, the standard deviation on $\Omega_{\rm m}$ and $S_8$ we report in \autoref{tab:avg_std_powerloss} is always lower than the error found by the DES collaboration \citep{DESY3-Secco2022++ModellingUncertainty} down to $\kmax \geq 0.10 \,h \rm Mpc^{-1}$. Of course our analysis is somewhat simplified since we do not consider real data and we are looking just at cosmological parameters. However, these results suggest promising prospects for LSST Y1 and highlight the importance of controlling the photo-$z$ bias.

For some combinations, such as BAHAMAS-Eagle, presented in the upper panel of \autoref{fig:corner-plots-Omegam-S8-DES-KiDS}, the different scale cuts also produce a shift on the mean value of $\Omega_{\rm m}$ and $S_8$. This is due to the different description of baryons in the two models: if the analysis model is stronger (weaker) than the one used to create the mock data it will bias the inference resulting in higher (lower) values of the $S_8$ parameter - intuitively, this effect is less severe for combination of models that describe baryons in a similar way (such as BAHAMAS and owlsAGN, see middle panels of \autoref{fig:corner-plots-Omegam-S8-DES-KiDS}).  
It is important to notice here that, as clearly visible in \autoref{fig:Pk-BaryonicEffects}, the Eagle simulation has almost the same effect as not considering any baryons in the range of scales we are interested in (hence the significant bias in the contour plot). 
This shift is more appreciable for cuts less stringent than $\kmax = 0.30\, h \rm Mpc^{-1}$, for which also the contour area remains almost the same.  

Despite these biases and weakened contraints, \autoref{fig:corner-plots-Omegam-S8-DES-KiDS} shows our first important result: even at very stringent scale cuts the tension with Planck measurements is still visible - especially with KiDS sized tension data, as expected since the underlying tension is larger.  
Note that the tension with Planck will increase or decrease depending on the chosen baryonic model. However, our analysis does not suggest a preferred model compared to another, but rather that with even minimal information on baryons (considering only the largest scales) we can still have good constraints and compatible results with previous surveys. 

Notice that, as expected, the cosmological parameters contours derived from DES and KiDS-like analyses exhibit similar behavior. The only notable difference, reflected in the shifted contours seen in \autoref{fig:corner-plots-Omegam-S8-DES-KiDS}, arises from the different underlying tension imposed by the respective fiducial values.

We remind the reader that, as explained in \autoref{Section:BaryonicFeedback}, so far we considered non-parametric models. The only exception is the case BAHAMAS-Bacco - this represents a different and less idealistic scenario. Predictably, while the behaviour at large scales is similar to previous cases (less affected by baryons), when including small scales - \textit{i.e.} larger $\kmax$ values - the constraining power on cosmological parameters decreases as we now vary six baryonic parameters. As a result, the contour area becomes nearly identical to that obtained without considering small scales, with $\kmax = 0.20\, h \rm Mpc^{-1}$.

\begin{table}
    \centering
    \begin{tabular}{|c|c|c|c|c|c|c|}
        \hline
        $\kmax$ & $\Omega_{\rm m}$ & $\sigma_8$ & $S_8$ \\
        \hline
        0.05 & 0.300 $\pm$ 0.022  &  0.768 $\pm$ 0.033  &  0.767 $\pm$ 0.029  \\
        0.10 & 0.301 $\pm$ 0.020  &  0.772 $\pm$ 0.029  &  0.772 $\pm$ 0.016  \\
        0.20 & 0.300 $\pm$ 0.019  &  0.775 $\pm$ 0.027  &  0.774 $\pm$ 0.001  \\
        0.30 & 0.299 $\pm$ 0.018  &  0.777 $\pm$ 0.025  &  0.775 $\pm$ 0.007  \\
        0.50 & 0.299 $\pm$ 0.018  &  0.778 $\pm$ 0.023  &  0.775 $\pm$ 0.006  \\
        0.80 & 0.299 $\pm$ 0.017  &  0.777 $\pm$ 0.022  &  0.775 $\pm$ 0.006  \\
        1.00 & 0.298 $\pm$ 0.016  &  0.778 $\pm$ 0.021  &  0.775 $\pm$ 0.006  \\ 
        \hline
    \end{tabular}
    \caption{Mean and standard deviation of cosmological parameters for each scale cut value.}
    \label{tab:avg_std_powerloss}
\end{table}

\begin{figure}
    \centering
    \includegraphics[scale=0.4]{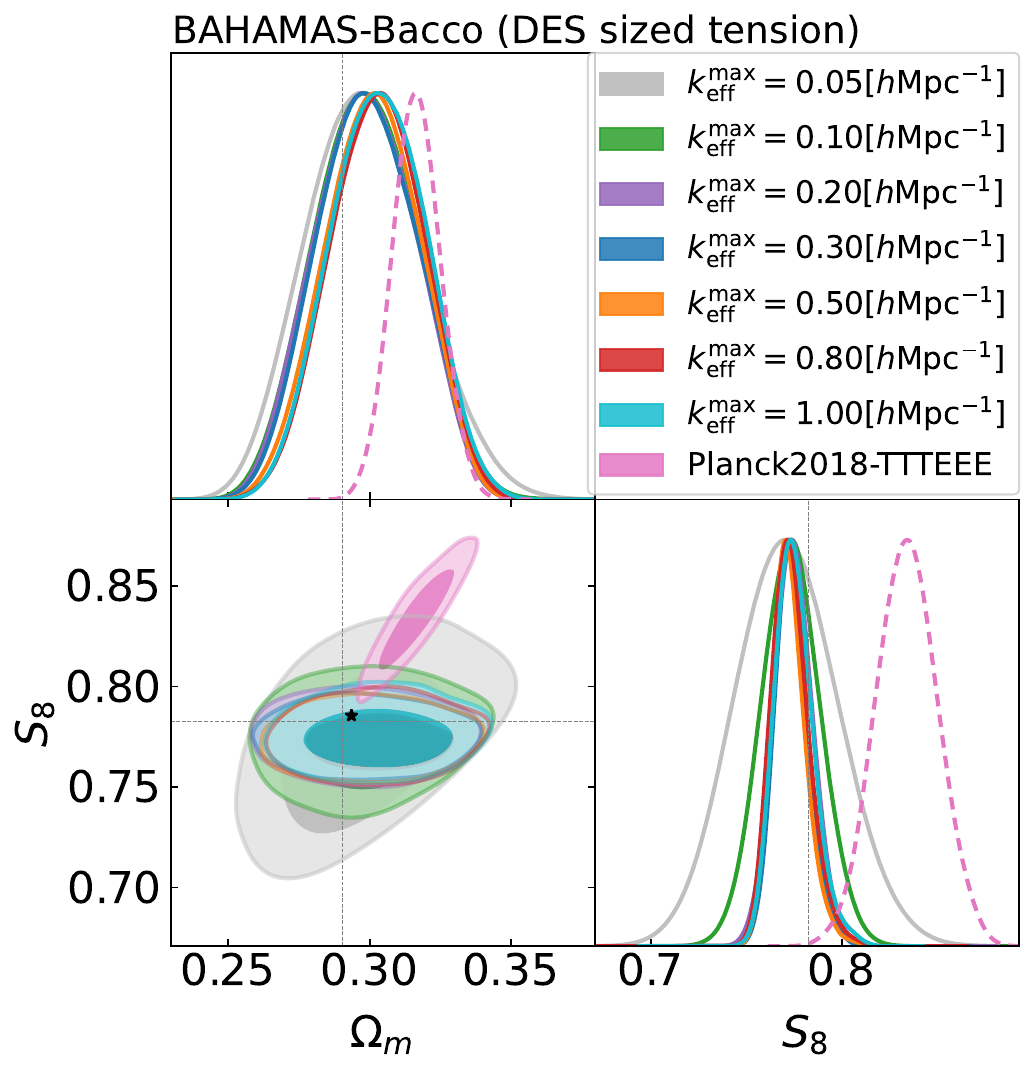}
    \caption{The $68\%$ and $95\%$ 2D confidence regions and 1D posteriors for $\Omega_{\rm m}$ and $S_8$ compared to the Planck contours (pink ellipses) for the case BAHAMAS-Bacco. Different choices of scale cuts, $\kmax$, are reported with different colours. The fiducial values used for creating the mock data (see \autoref{tab:fiducials}) are shown with the dashed grey lines. The black ${*}$ symbols show the maximum a posteriori (MAP) estimate at $\kmax=0.10 \,h \rm Mpc^{-1}$. Note that some of the sampled data have been rejected because they lie outside the parameter range of Bacco, resulting in the flattening of the contours, especially visible at $\kmax=0.05\, h\rm Mpc^{-1}$.}
    \label{fig:corner-trueDES-bahamas-bacco}
\end{figure}

\subsection{Quantifying the $S_8$ tension}
\begin{figure*}
    \centering
    \includegraphics[width=\textwidth]{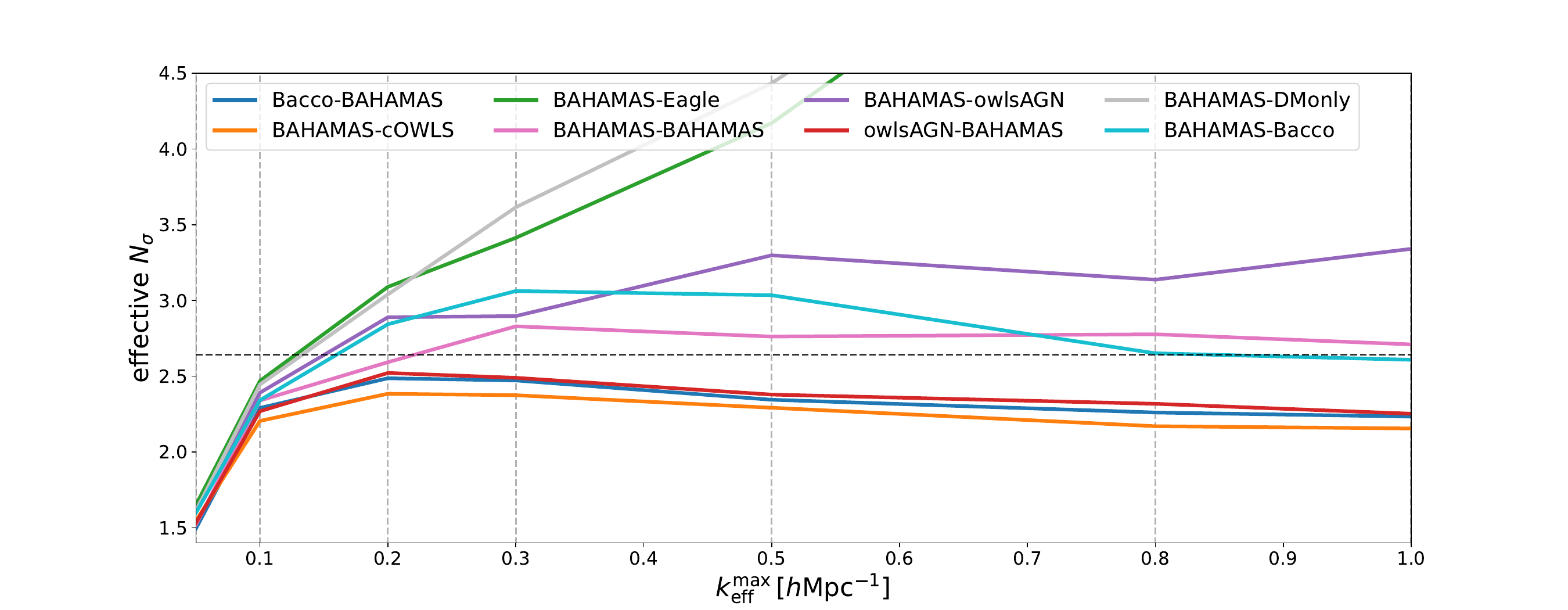} 
    \caption{Effective number of sigma at different $\kmax$ values for all model combination obtained with the tension metric ``parameter difference''. Mock data build from DES Y3 like fiducial values. The horizontal dotted line represents the true $N_\sigma$ (\textit{i.e.} the expected value between the Planck results and the trivial case BAHAMAS-BAHAMAS), $N^{\rm true}_\sigma \simeq 2.6$.}
    \label{fig:measured-tension-vs-kmax-all}
\end{figure*}

As previously described, we have used the \texttt{tensiometer} code to compute the $S_8$ tension between all cases listed in \autoref{tab:different-model-combinations} and the Planck measurements. The results are reported in \autoref{fig:measured-tension-vs-kmax-all}, where the measured tension is shown as function of the used scale cut. The horizontal black dashed line in the figure represents the ``true tension'' $N^{\rm true}_\sigma \simeq 2.6$, measured between Planck measurements and the run with BAHAMAS-BAHAMAS as baryonic models with no scale cuts applied.

All model combinations yield values similar to the true one, indicating that the choice of baryonic model does not have a significant impact on the measured tension.
The only exceptions are the results for the unphysical BAHAMAS-DM only case and BAHAMAS-Eagle - which, recalling \autoref{fig:Pk-BaryonicEffects}, behaves very similarly to the DM only scenario in this range of scales. The measured $N_\sigma$ has the same scale cut dependency for all cases excluding the two outliers; each measured tension stays within the $25\%$ of the true value (considering the range $\kmax = 0.10 - 1.0\,h\rm Mpc^{-1}$).

As expected, the measured tension decreases as the scale cuts become more stringent, and for $\kmax<0.10\,h\rm Mpc^{-1}$, the error bars become too large causing the tension to diminish. At $\kmax=0.20\,h\rm Mpc^{-1}$ all cases converge around the true tension, with a difference of approximately $17\%$ from $N_\sigma^{\rm true}$.
The bias described in the previous section is seen also here and again, for scale cuts higher than $\kmax=0.30\,h\rm Mpc^{-1}$, there is no significant change in the measured tension of each chain.

We can therefore state that for scale cuts smaller than $\kmax=0.20\,h\rm Mpc^{-1}$ the effect of baryons is negligible. However, as we impose stricter scale cuts, we experience a significant loss in constraining power, which limits our ability to draw meaningful conclusions.

In the case of a KiDS-like survey, we find a nominal $N_\sigma\simeq5$ at $\kmax = 0.10\,h \rm Mpc^{-1}$, but because of the prior boundary effect on $\Omega_{\rm m}$ (see \autoref{fig:corner-plots-Omegam-S8-DES-KiDS} caption) this is likely to be an underestimate. As expected the tension becomes smaller with smaller $\kmax$, while at higher $\kmax$ the \texttt{tensiometer} metrics diverges as the shift probability is 1.
This implies that LSST Y1 observations with $\kmax \sim 0.1 \,h \rm Mpc^{-1}$ with an underlying tension similar to KiDS will show a tension unambiguously, regardless of whether an incorrect model for baryons is assumed.

%

\subsection{Prior checks}
\begin{figure*}
    \centering
    \includegraphics[width=\textwidth]{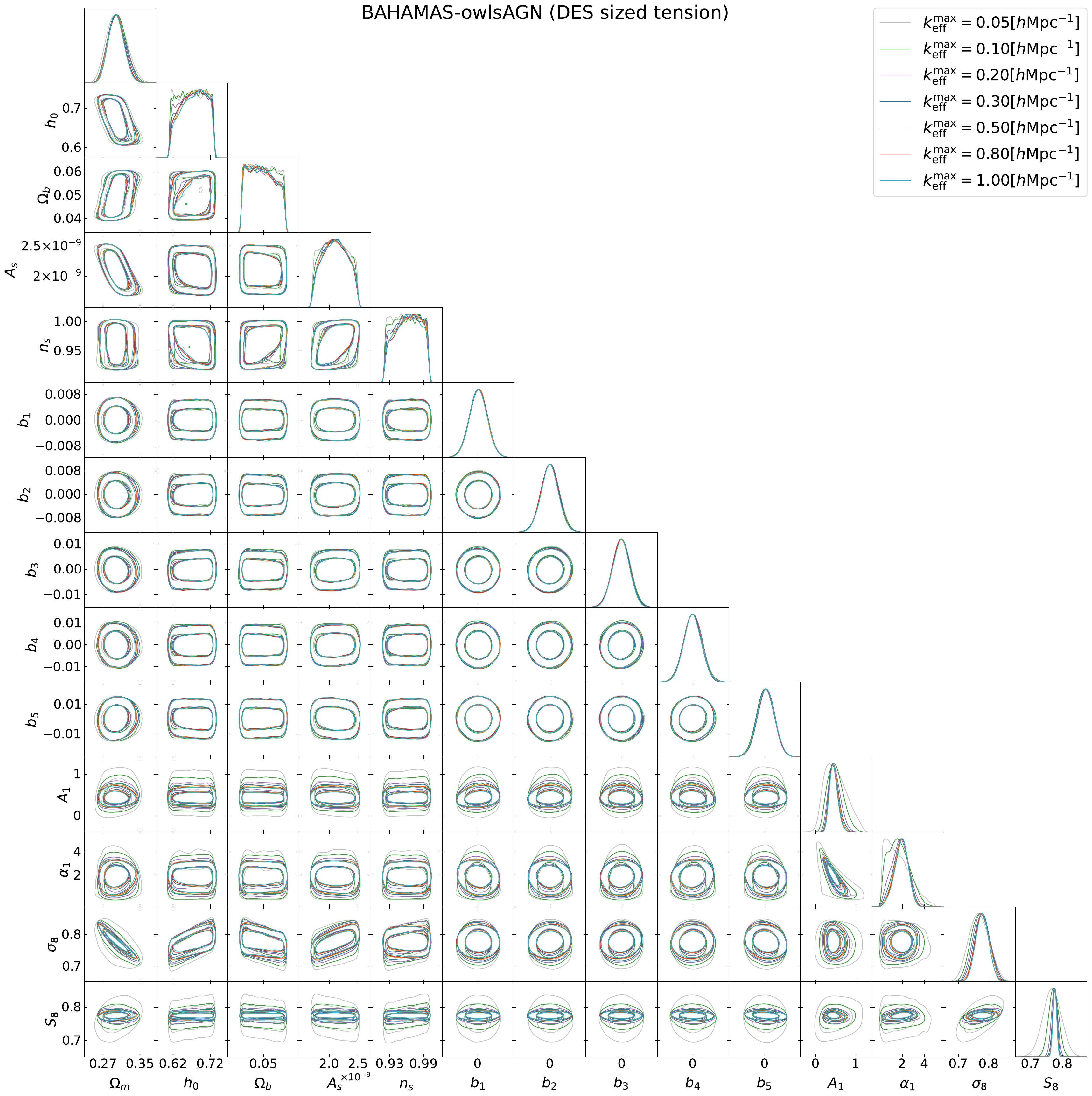}
    \caption{Contour plot for all parameters for the case BAHAMAS-owlsAGN.}
    \label{fig:cornerplot-all-params}
\end{figure*}

To ensure that all parameters are well constrained within their prior ranges, we run \cosmosis with the \texttt{Apriori} sampler, which draws samples from the prior and evaluates their likelihood. As expected, we find that $h_0$, $\Omega_{\rm b}$ and $n_{\rm s}$ remain unconstrained (see \autoref{fig:cornerplot-all-params}), since weak lensing is not effective at constraining them (this behaviour is also seen in other papers with broader priors, \textit{eg.} \citet{DESY3-Secco2022++ModellingUncertainty}, and explained in \citet{Hall2021}). Similarly, the five bias parameters $\delta_z^i$ are prior dominated. However, $\sigma_8$, $S_8$ and $\Omega_{\rm m}$ are well constrained.

\subsection{Requirements on number density distribution} \label{sec:nz_requirements}
\begin{figure*}
    \centering
    \includegraphics[scale=0.42]{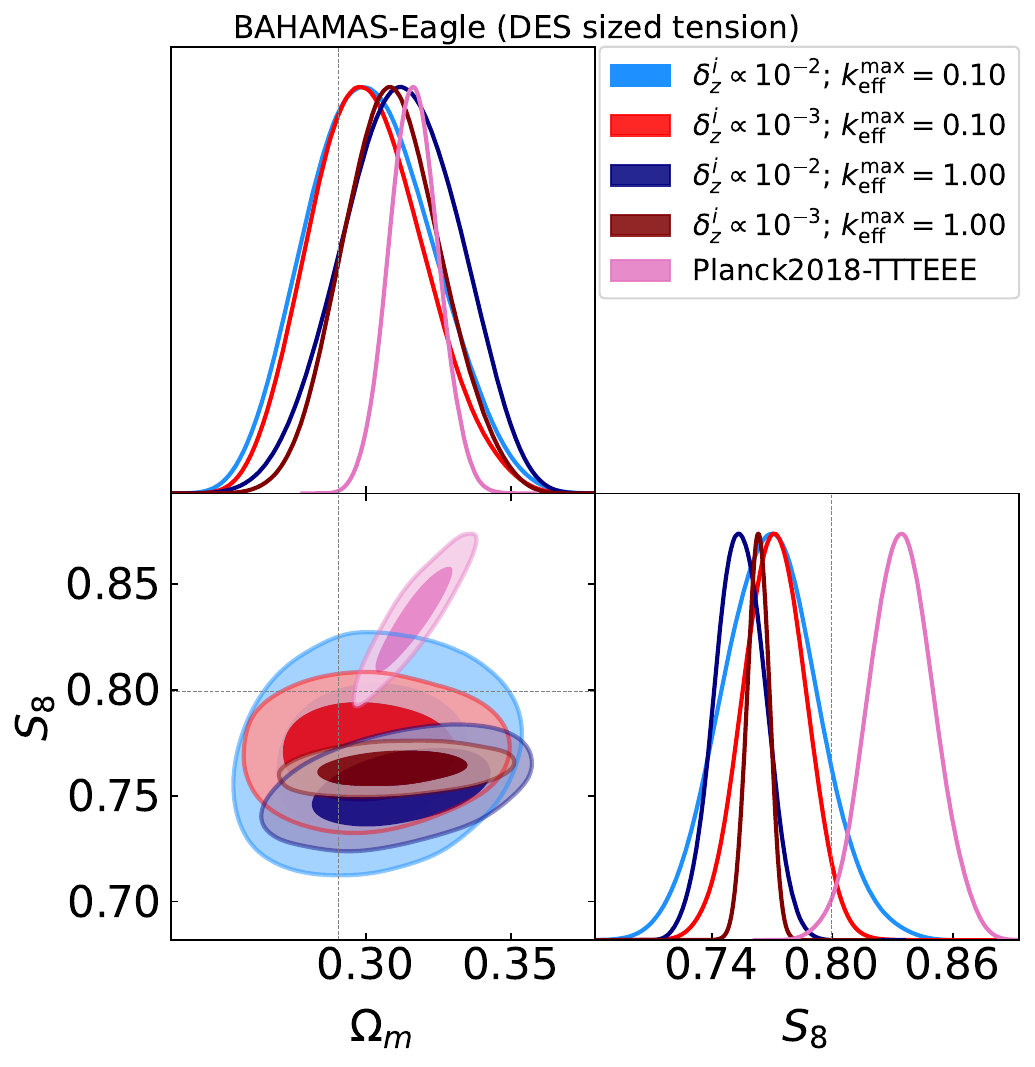} \quad \includegraphics[scale=0.42]{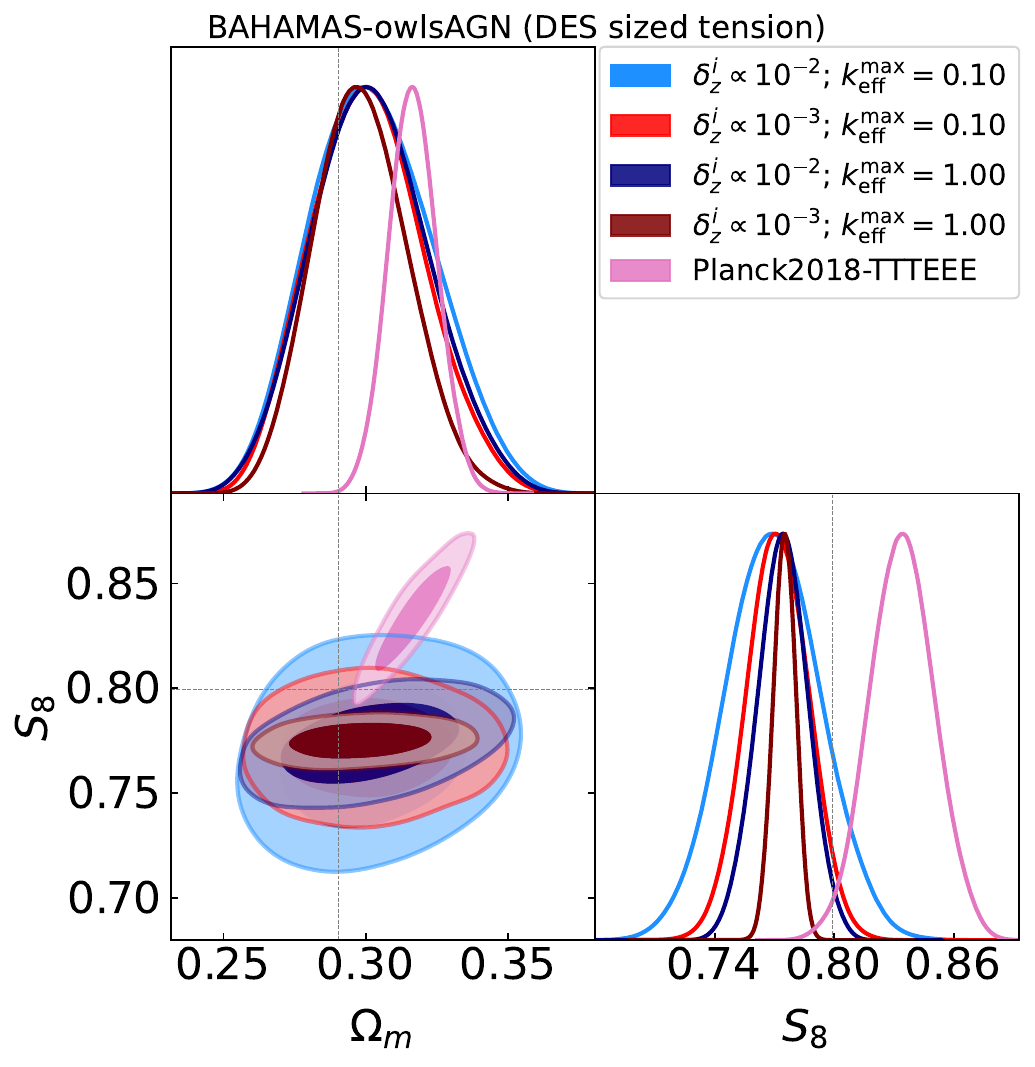}
    \caption{The $68\%$ and $95\%$ 2D confidence regions and 1D posteriors for $\Omega_{\rm m}$ and $S_8$ for the cases BAHAMAS-Eagle (left) and BAHAMAS-owlsAGN (right). The values of $\kmax$ are measured in $h\rm Mpc^{-1}$. The red contours are the result of the analysis with biases $\delta_z^i$ as in the \citetalias{DESC-SRD-2018} (see \autoref{tab:priors}), while in blue is reported the same analysis but with priors of $\delta_z^i$ incremented by a factor of 10. Darker colors correspond to $\kmax=1.00\,h\rm Mpc^{-1}$, while lighter colors to $\kmax=0.20\,h\rm Mpc^{-1}$.} 
    \label{fig:bigger-bias-bahamas-eagle-owls}
\end{figure*}

The standard deviation of the photo-$z$ bias prior has been set to approximately $10^{-3}$ as in \citetalias{DESC-SRD-2018}, but, this might be overly optimistic. To explore this further, we consider two model combinations as tests: BAHAMAS-Eagle and BAHAMAS-owlsAGN with the DES sized tension. These correspond to the two cases where we have less and more agreement between the mock data and the analysis model.
We run again the same pipeline, with the standard deviation of the $\delta_z^i$ priors listed in \autoref{tab:priors} increased by an order of magnitude. 
As expected (see \autoref{fig:bigger-bias-bahamas-eagle-owls}), since this value represents large uncertainty in the redshift distribution $n(z)$, it results in broader contours for the cosmological parameters. Consequently, the measured tension with Planck becomes less significant. For instance, in the case of an analysis using DES-like fiducials, where the tension with Planck is already lower, the resulting tension is below the true value $N_\sigma^{\rm true}$ for $\kmax \lesssim 0.50\,h\rm Mpc^{-1}$, and remains consistently below 1.8 (\textit{i.e.,} indicating no statistically significant tension) in the BAHAMAS-cOWLS case.

This result holds a significant implication: the baryon-free large scale tension detection we have studied in this work can be seen only if LSST Y1 achieves measuring the redshift distributions with high accuracy ($\delta_z^i \propto 10^{-3}$). Otherwise, we will lose constraining power, which directly affects tension measurements, meaning that we would not be able to distinguish the effects of the baryons (and their impact on the $S_8$ tension) even with a stringent scale cut of $\kmax=0.20\,h\rm Mpc^{-1}$.

\section{Conclusions}\label{Sec:Conclusions}
Modelling baryonic feedback is crucial for Stage IV surveys, as their high sensitivity makes them even more vulnerable to producing biased results if such effects are not properly accounted for.
However, due to their complex physics, accurately modeling these effects remains a challenge, and we must rely on hydrodynamical simulations. It is therefore important to assess how significantly future data will be affected by baryonic processes and whether it is statistically meaningful to use only the unaffected large scales.

Although the nature of the $S_8$ tension is an open question - and it is still unclear whether it is a real, physical tension - recent studies suggest that non-linear physics may play an important role in solving this discrepancy \citep{PrestonAmonEfstathiou_S8tension2023}, while others claim that baryonic feedback alone would not be sufficient \citep{Flamingo2023McCarthy, Salcido&McCarthy2024-baryonS8tension, 2024arXiv240320323T}. Therefore, understanding these effects is crucial.

In this work, we have studied the impact of baryonic feedback on small scales in an LSST Y1-like lensing survey and and assessed the information that can be extracted from the survey data when applying (stringent) scale cuts to discard small scales. 
We adopted a more physically motivated method to define the cuts, accounting for their redshift dependence, which were then applied to generated data chains using the \textsc{nautilus} sampler. Various combinations of baryonic models were employed to generate mock data and to perform our analysis, aiming to be agnostic about the true model.

The loss of constraining power due to scale cuts mainly affects the $\sigma_8$ and $S_8$ parameters, leaving the others mostly unchanged. More specifically, while the error on $\Omega_{\rm m}$ slightly increases with stricter scale cuts, the error on $S_8$ increases up to five times in the extreme case with $\kmax=0.05\,h\rm Mpc^{-1}$. Nonetheless, the constraints are better or comparable to those achieved by DES.

We have shown that considering different baryonic feedback models causes shifts in the inferred values of $\Omega_{\rm m}$ and $S_8$, depending on how closely the analysis model matches the underlying physics of the mock data.

Moreover, while baryons affect the data for $k\gtrsim0.20\,h\rm Mpc^{-1}$,
we can clearly distinguish between our fiducial measurements and Planck results even considering $\kmax=0.10\,h\rm Mpc^{-1}$, particularly for datasets with higher underlying tension (KiDS-like). 
In general, we found that, excluding the cases involving Eagle and DM-only as analysis model (\textit{i.e.} the most extreme cases), the choice of baryonic model does not strongly influence the measured tension with Planck. Specifically, all measured tensions remain within $25\%$ of the underlying imposed tension across all scales, and within the $15\%$ at scales not affected by baryons. This suggests that even with minimal information on baryons, useful constraints can still be derived from large-scale data, yielding results compatible with previous surveys. 

Finally, we emphasize that the conclusions of this study rely on LSST's ability to achieve high-precision measurements of the redshift distribution.
By testing more relaxed priors on $\delta_z^i$, it was found that increasing them by an order of magnitude leads to significantly broader parameter contours, reducing the measured tension with Planck. This implies that unless LSST achieves a high precision in redshift distribution measurements (with uncertainties as low as $\delta_z^i \propto 10^{-3}$), the ability to constrain cosmological parameters and assess the impact of baryonic feedback - particularly regarding the $S_8$ tension - will be compromised.

It is important to acknowledge that our analysis is optimistic, as we focused solely on cosmological parameters while neglecting other relevant complexities, such as intrinsic alignments and varying baryonic parameters (these were fixed for all but one of our analyses). 
 
So far, we have been working with weak gravitational lensing, however, more powerful statistics can come from the combination with clustering in the so called ``$3\times2$pt'' \citep{PratZuntz-3x2ForecastRubin2023}. Since we are focusing on large-scales only, using a linear bias model for clustering could be sufficient. Incorporating such an analysis, along with different intrinsic alignment models and varied baryonic parameters, would provide a more comprehensive picture of baryonic feedback and its impact on cosmological constraints.

\begin{figure*}
    \centering
    \includegraphics[scale=0.42]{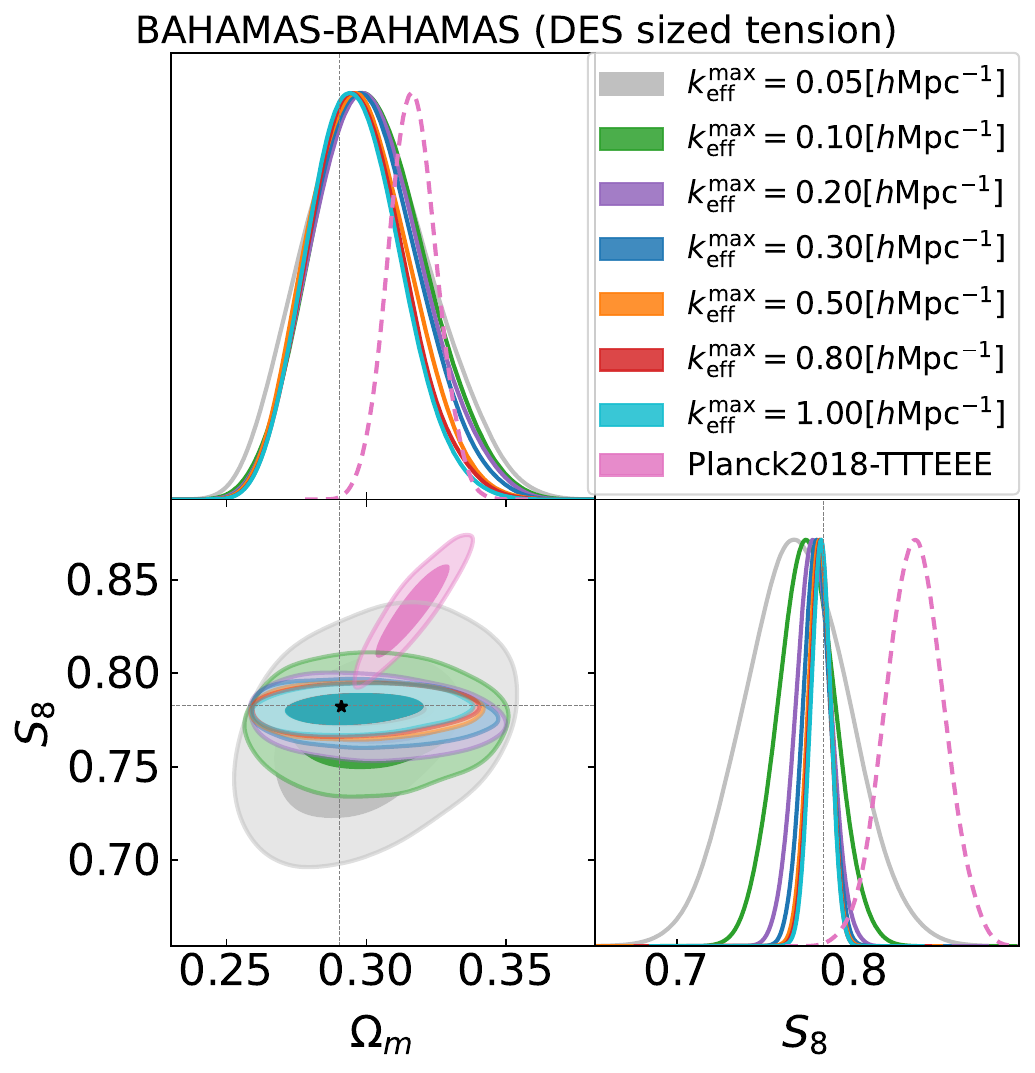} \quad \includegraphics[scale=0.42]{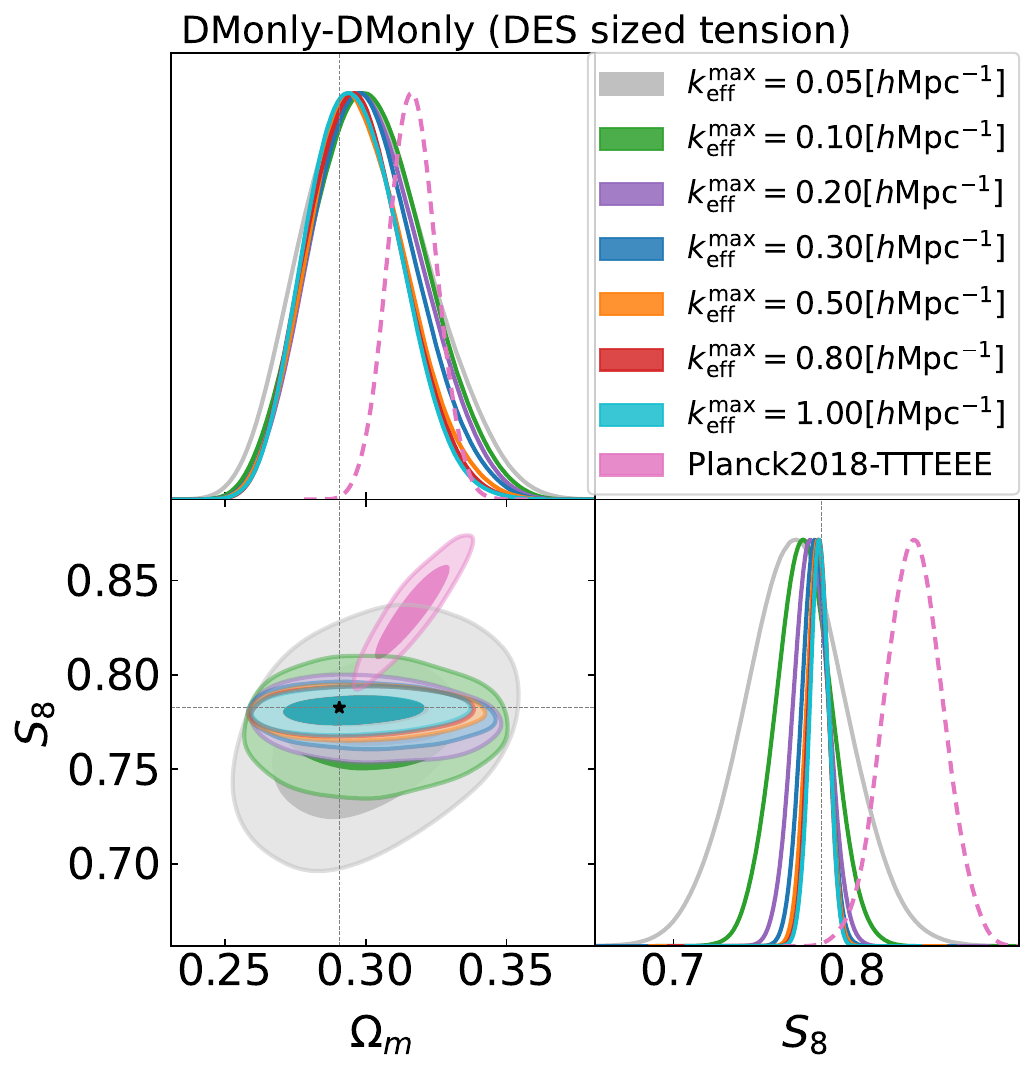}
    \caption{The $68\%$ and $95\%$ 2D confidence regions and 1D posteriors for $\Omega_{\rm m}$ and $S_8$ for the cases BAHAMAS-BAHAMAS (left) and DMonly-Dmonly (right). The values of $\kmax$ are measured in $h\rm Mpc^{-1}$. The black ${*}$ symbols show the maximum a posteriori (MAP) estimate at $\kmax=0.10 \,h \rm Mpc^{-1}$.} 
    \label{fig:corner-bahamas-dmonly}
\end{figure*}

\appendix
In addition to the scenarios described throughout the paper, we run two other cases as validation tests: BAHAMAS-BAHAMAS and DMonly-DMonly. The resulting corner plots are shown in \autoref{fig:corner-bahamas-dmonly}, from which we can draw the same conclusions on loss of constraining power and tension detectability as presented in \autoref{SubSec:Constrains}. Moreover, we can confirm that the shift seen in \autoref{fig:corner-plots-Omegam-S8-DES-KiDS} is primarily due to the different baryonic model assumed in the mock data and in the analysis model, while the small residual offset with the fiducial values at small scales can be identified as a projection effect \citep{projeff-Hadzhiyska2020, projeff-GomezValent2022}, as the effect of baryons is negligible at these scales.

\section*{Acknowledgements}
OT is funded by an Science and Technology Facilities Council (STFC) studentship. AP is a UK Research and Innovation Future Leaders Fellow [grant MR/X005399/1]. For the purpose of open access, the author has applied a Creative Commons Attribution (CC BY) licence to any Author Accepted Manuscript version arising from this submission.

\bibliographystyle{mnras_2author}
\bibliography{references.bib}

\end{document}

%% file: authorlist.tex
\email[$^\star$: Corresponding author: ]{ottavia.truttero@ed.ac.uk}

\author{Ottavia~Truttero$^{\star1}$}
\author{Joe~Zuntz$^{1}$}
\author{Alkistis~Pourtsidou$^{1,2}$}
\author{Naomi~Robertson$^{1}$}

\affiliation{$^{1}$ Institute for Astronomy, University of Edinburgh, Royal Observatory, Blackford Hill, Edinburgh, EH9 3HJ, U.K.}
\affiliation{$^{2}$Higgs Centre for Theoretical Physics, School of Physics and Astronomy, University of Edinburgh, Edinburgh EH9 3FD, U.K.}